\documentclass[10pt]{article}
\usepackage{graphicx}
\usepackage{amsmath}
\usepackage{amssymb}
\usepackage{balance}
\usepackage{bm}
\usepackage{caption2}
\usepackage{indentfirst}

\begin{document}
\bibliographystyle{prsty}
\begin{center}
{\large {\bf \sc{ The scalar and pseudoscalar hidden-charm tetraquark states with QCD sum rules   }}} \\[2mm]
Zun-Yan Di$^{1,2}$, Zhi-Gang Wang$^{1}$\footnote{E-mail: zgwang@aliyun.com. }    \\
$^{1}$ Department of Physics, North China Electric Power University, Baoding 071003, P. R. China \\
$^{2}$ School of Nuclear Science and Engineering, North China Electric Power University, Beijing 102206, P. R. China
\end{center}

\begin{abstract}
Based on the diquark configuration, we construct the diquark-antidiquark interpolating
tetraquark currents with $J^{PC}=1^{-\pm}$ and $1^{+\pm}$,
which can couple to the scalar and pseudoscalar tetraquark states respectively,
since they are not conserved currents.
Then we investigate their two-point correlation functions including the contributions of the vacuum
condensates up to dimension-10 and extract the masses and pole residues of the tetraquark states
with $J^{PC}=0^{+\pm}$ and $0^{-\pm}$ through the QCD sum rule approach.
The predicted masses can be confronted with the experimental data in the future.
Moreover, we briefly discuss the possible decay patterns of the tetraquark states.
\end{abstract}

 PACS number: 12.39.Mk, 12.38.Lg

Key words: Hidden-charm tetraquark state,  QCD sum rules

\section{Introduction}
Since the discovery of the $X(3872)$ resonance by Belle collaboration in 2003 \cite{X(3872)},
more and more exotic hadrons have been observed and confirmed experimentally,
such as the charmonium-like XYZ states \cite{XYZ}, hidden-charm pentaquarks \cite{Pc}, etc.
These resonances with four or five valence quarks cannot be interpreted as conventional quark-antiquark mesons
or three-quark baryons in the quark model \cite{Quark-Model}.
They are new blocks of QCD matter,
which provide an important
platform to deepen our understanding of the low energy behaviors of QCD.

Facing the large amount of data on the exotic states, the theoretical researchers of high energy physics
have proposed several models to explain their nature,
such as molecule, multiquark state, hadrocharmonium, hybrids, kinematical effects, etc.
In the molecular picture,
a tetraquark (pentaquark) state is explained as a hadronic molecule of two mesons (one meson and one baryon)
\cite{Hadron-Molecules,Molecular-Charmonium}.
The multiquark state interpretation is based on the phenomenological diquark picture \cite{Diquarks},
in which a tetraquark state is assumed to be a diquark-antidiquark object [8-19]
and a pentaquark state is a diquark-diquark-antiquark object [20-29],
bound by gluonic exchanges.
In the hadroquarkonium picture for multiquark exotics,
the heavy-quark pair $Q\bar{Q}$ forms a compact core
about which the light $q\bar{q}$ or $qqq$ forms a quantum-mechanical cloud \cite{Quarkonium,Hadro-Charmonium}.
Here, we simply introduce the above three popular models.
For more reviews of the theoretical interpretations, see Ref.\cite{review-1,review-2}.
Unfortunately, as so far, no single model naturally accommodates all the observed states.
It will be a long way to reveal the nature of the multiquark candidates completely.

In addition, the observations of these exotic states stimulate the arguments for more possible multiquark states.
In Ref.\cite{article-Di}, we have studied the possible  scalar hidden-charm $cu\bar{c}\bar{d}$ $(cu\bar{c}\bar{s})$
tetraquark states by constructing the corresponding $C\otimes C$
and $C\gamma_\mu \gamma_5\otimes \gamma_5 \gamma^\mu C$
type scalar interpolating currents.
In this article, we investigate the other possible scalar
and pseudo-scalar hidden-charm tetraquark states with different structures.
Specifically, we construct the $C\otimes \gamma_\mu C$
and $C \gamma_5\otimes \gamma_5 \gamma_\mu C$ type
interpolating tetraquark currents with $J^{PC}=1^{-\pm}$ and $1^{+\pm}$
in the diquark configuration,
calculate their two-point correlation functions,
and extact the spectral densities
for the scalar and pseudoscalar tetraquark states
through the tensor analysis method.
Then we perform the QCD sum rule analysis and
obtain the masses and pole residues of the hidden-charm tetraquark states with $J^{PC}=0^{+\pm}$ and $0^{-\pm}$.

This article is organized as follows.
In section 2, we construct the vector and axial-vector interpolating tetraquark currents,
extact the spectral densities for the scalar and pseudoscalar tetraquark states up to dimension-10
and derive the masses and pole residues of the scalar and pseudoscalar tetraquark states with the QCD sum rules.
The numerical results and discussions are performed in section 3.
The last section is reserved for our conclusion.

\section{QCD sum rules for the $J^{PC}=0^{+\pm}$ and $0^{-\pm}$  hidden-charm tetraquark states}
To begin, we construct the diquark-antidiquark interpolating
tetraquark currents with $J^{PC}=1^{-\pm}$ and $1^{+\pm}$, based on the diquark configuration.
The vector and axial-vector interpolating currents are
\begin{eqnarray}
J_{\mu }^{\,t,a}(x)&=&\frac{\epsilon^{i j k}\epsilon^{i m n}}{\sqrt{2}}\left\{ u^{j T} (x)Cc^{k} (x) \bar{d}^{m}(x)\gamma_{\mu} C \bar{c}^{n T}(x)+t u^{j T} (x)C\gamma_{\mu}c^{k} (x) \bar{d}^{m}(x) C \bar{c}^{n T}(x)\right\}
\end{eqnarray}
and
\begin{eqnarray}
J_{\mu}^{\,t,b}(x)&=&\frac{\epsilon^{i j k}\epsilon^{i m n}}{\sqrt{2}}\left\{ u^{j T} (x)C\gamma_{5}c^{k} (x) \bar{d}^{m}(x)\gamma_{\mu} C \bar{c}^{n T}(x)\right. \nonumber\\
&&\left.+t u^{j T} (x)C\gamma_{\mu}c^{k} (x) \bar{d}^{m}(x)\gamma_{5} C \bar{c}^{n T}(x)\right\}
\end{eqnarray}
respectively, where
the $i$, $j$, $k$, $m$ and $n$ are color indexes, the $C$ is the
charge conjugation matrix.
Under charge conjugation transform $\widehat{C}$,
the currents $J_{\mu}^{\,t,a/b}(x)$ have the
properties,
\begin{eqnarray}
\widehat{C}J_{\mu }^{\,t,a/b}(x)\widehat{C}^{-1}&=&\pm J_{\mu }^{\,t,a/b}(x)\mid_{u\leftrightarrow d}\,\,\ \text{for} \,\,\ t=\pm \ ,
\end{eqnarray}
which originate from the charge conjugation properties of the scalar, pseudoscalar and axial-vector diquark
states,
\begin{eqnarray}
\widehat{C}\left[\epsilon^{i j k}q^{j}C\gamma_5 c^{k}\right]\widehat{C}^{-1}&=&\epsilon^{i j k}\bar{q}^{j}\gamma_5C \bar{c}^{k}\, ,\nonumber\\
\widehat{C}\left[\epsilon^{i j k}q^{j}C c^{k}\right]\widehat{C}^{-1}&=&\epsilon^{i j k}\bar{q}^{j}C \bar{c}^{k}\, ,\nonumber\\
\widehat{C}\left[\epsilon^{i j k}q^{j}C\gamma_\mu c^{k}\right]\widehat{C}^{-1}&=&\epsilon^{i j k}\bar{q}^{j}\gamma_\mu C \bar{c}^{k}\, ,
\end{eqnarray}
where $q=u$, $d$.
Thus the superscript $t=\pm$ of the interpolating currents $J_{\mu }^{\,t,a/b}(x)$
can correspond
the positive and negative charge conjugations for the vector and axial-vector tetraquark states.

In the following, we compute
\begin{eqnarray}
p^{\mu}J_{\mu }^{\,t,a}(x)&=&i\partial^{\mu}J_{\mu }^{\,t,a}(x) \nonumber\\
&=&-\frac{\epsilon^{i j k}\epsilon^{i m n}}{\sqrt{2}}\left(m_c-m_q\right)\left\{ u^{j T} (x)Cc^{k} (x) \bar{d}^{m}(x) C \bar{c}^{n T}(x)  \right.  \nonumber\\
&&\left.-t u^{j T} (x)C c^{k} (x) \bar{d}^{m}(x) C \bar{c}^{n T}(x)\right\}+\cdots \, ,\label{scalar}\\
p^{\mu}J_{\mu}^{\,t,b}(x)&=&i\partial^{\mu}J_{\mu}^{\,t,b}(x) \nonumber\\
&=&-\frac{\epsilon^{i j k}\epsilon^{i m n}}{\sqrt{2}}\left(m_c-m_q\right)\left\{ u^{j T} (x)C\gamma_{5}c^{k} (x) \bar{d}^{m}(x) C \bar{c}^{n T}(x)   \right.  \nonumber\\
&&\left.-t u^{j T} (x)C c^{k} (x) \bar{d}^{m}(x)\gamma_{5} C \bar{c}^{n T}(x)\right\}+\cdots \, ,\label{pseudoscalar}
\end{eqnarray}
where $m_q=m_u=m_d$, the $p^\mu$ is the momentum of the current,
which is equivalent to the sum of the constituent quarks' momenta
and can be replaced by the $i\partial^\mu$ in momentum space.
In Eqs.\eqref{scalar}-\eqref{pseudoscalar}, the derivative operator acts on all
quark fields including $u^{jT}(x)Cc^{k}(x)$
and the relevant terms are not written out because they are complicated, which can not
be simplified by the Dirac equation.
From the Eqs.\eqref{scalar}-\eqref{pseudoscalar}, we can see that
the $\partial^{\mu}J_{\mu }^{\,t,a/b}(x)\neq0$,
hence the currents $J_{\mu }^{\,t,a/b}(x)$ are not conserved
and can couple to the scalar and pseudoscalar tetraquark states, respectively.
Besides, the Eqs.\eqref{scalar}-\eqref{pseudoscalar} indicate that
the superscript $t=\mp$ of the interpolating currents $J_{\mu }^{\,t,a/b}(x)$
can correspond
the positive and negative charge conjugations for the scalar and pseudoscalar tetraquark states.

The two-point correlation functions of the vector and axial-vector currents are written as
\begin{eqnarray}\label{correlation-function}
\Pi_{\mu \nu}^{t,a/b}\left(p\right)&=&i\int d^{4} x e^{ip\cdot x}
\langle0|T\left\{J_{\mu}^{\,t,a/b}(x) J_{\nu}^{{\,t,a/b}\,\dag}(0)\right\}|0\rangle\, \nonumber\\
&=&\Pi_{1}^{t,a/b}\left(p\right)\left(-g_{\mu \nu}+\frac{p_{\mu}p_{\nu}}{p^2}\right)+\Pi_{0}^{t,a/b}\left(p\right)p_{\mu}p_{\nu}\, .
\end{eqnarray}
There are two parts of $\Pi_{\mu \nu}^{t,a/b}\left(p\right)$ with different Lorentz structures
because the currents $J_{\mu}^{\,t,a/b}(x)$ are not conserved currents.
$\Pi_{1}^{t,a/b}\left(p\right)$ are related
to the vector and axial-vector tetraquark states,
while $\Pi_{0}^{t,a/b}\left(p\right)$ are the scalar and pseudoscalar current polarization functions.

At the phenomenological side,
we insert a complete set of intermediate hadronic states
with the same quantum numbers as the current operators $J_{\mu}^{\,t,a/b}(x)$
into the correlation functions $\Pi_{\mu \nu}^{t,a/b}\left(p\right)$
to obtain the hadronic representation.
After isolating the ground state contributions of the
hidden-charm tetraquark states from the pole terms,
we get the following results,
\begin{eqnarray}\label{correlation-function}
\Pi_{\mu \nu}^{t,a/b}\left(p\right)
&=&\frac{\lambda_{Z_{1}^{\,t,a/b}}^2}{M_{Z_{1}^{\,t,a/b}}^2-p^2}\left(-g_{\mu \nu}+\frac{p_{\mu}p_{\nu}}{p^2}\right)+\frac{\lambda_{Z_{0}^{\,t,a/b}}^2}{M_{Z_{0}^{\,t,a/b}}^2-p^2}p_{\mu}p_{\nu}+\cdots\, ,
\end{eqnarray}
where the pole residues $\lambda_{Z_{1}^{\,t,a/b}}$ and $\lambda_{Z_{0}^{\,t,a/b}}$ are defined by
\begin{eqnarray}
\langle0|J_{\mu}^{\,t,a/b}(0)|Z_{1}^{\,t,a/b}(p)\rangle &=& \lambda_{Z_{1}^{\,t,a/b}}\varepsilon_\mu\ ,\nonumber\\
\langle0|J_{\mu}^{\,t,a/b}(0)|Z_{0}^{\,t,a/b}(p)\rangle &=& \lambda_{Z_{0}^{\,t,a/b}}p_\mu\ ,
\end{eqnarray}
the $Z_{1}^{\,t,a/b}$and $Z_{0}^{\,t,a/b}$
are the ground states of the spin-1 and spin-0
hidden-charm tetraquark states, respectively,
and the $\varepsilon_\mu$ are the polarization vectors of
the vector and axialvector tetraquark states.
In Refs.\cite{vector,axialvector}, the authors have
chosen the tensor structure $-g_{\mu \nu}+\frac{p_{\mu}p_{\nu}}{p^2}$ for analysis
and investigated
the corresponding vector and axial-vector tetraquark states $Z_{1}^{\,t,a/b}$.
In this article, we make Eq.\eqref{correlation-function} multiplied by the
$p^\mu$,
\begin{eqnarray}
p^{\mu}\Pi_{\mu \nu}^{t,a/b}\left(p\right)&=&\frac{\lambda_{Z_{1}^{\,t,a/b}}^2}{M_{Z_{1}^{\,t,a/b}}^2-p^2}\left(-g_{\mu \nu}p^{\mu}+\frac{p^\mu p_{\mu}p_{\nu}}{p^2}\right)
+\frac{\lambda_{Z_{0}^{\,t,a/b}}^2}{M_{Z_{0}^{\,t,a/b}}^2-p^2}p_{\mu}p^\mu p_{\nu}+\cdots\, ,
\end{eqnarray}
to eliminate contaminations of the vector and axial-vector tetraquark states,
and study the remaining scalar and pseudoscalar tetraquark states $Z_{0}^{\,t,a/b}$.

Now, we take a short digression to study the contributions of the intermediate meson-loops
to the correlation function $\Pi_0^{-,a}(p)$ for the current $J_\mu^{-,a}(x)$ as an example,
the current $J_\mu^{-,a}(x)$ has non-vanishing couplings with the scattering states
$\eta_c\pi^+$, $J/\psi\rho^+(770)$, $\bar{D}^0D^+$, etc.
\begin{eqnarray}
\Pi_0^{-,a}\left(p\right)
&=&-\frac{\widehat{\lambda}_{Z_{0}^{\,-,a}}^2}{p^2-\widehat{M}_{Z_{0}^{\,-,a}}^2-\Sigma_{\eta_c\pi^+}(p)
-\Sigma_{J/\psi\rho^+(770)}(p)-\Sigma_{\bar{D}^0D^+}(p)+\ldots}+\cdots\ ,
\end{eqnarray}
where the $\widehat{\lambda}_{Z_{0}^{\,-,a}}$ and $\widehat{M}_{Z_{0}^{\,-,a}}$
are bare quantities to absorb the divergences in the self-energies
$\Sigma_{\eta_c\pi^+}(p)$, $\Sigma_{J/\psi\rho^+(770)}(p)$, $\Sigma_{\bar{D}^0D^+}(p)$, etc.
The renormalized self-energies contribute a finite imaginary part to modify the dispersion relation,
\begin{eqnarray}
\Pi_0^{-,a}\left(p\right)
&=&-\frac{\lambda_{Z_{0}^{\,-,a}}^2}{p^2-M_{Z_{0}^{\,-,a}}^2+i\sqrt{p^2}\Gamma(p^2)}+\cdots\, .
\end{eqnarray}
In previous works, we observed that the effects of the finite widths, such as
$\Gamma_{X(4500)}=92\pm21_{-20}^{+21}\ \text{MeV}$,
$\Gamma_{X(4700)}=120\pm31_{-33}^{+42}\ \text{MeV}$,
$\Gamma_{Z_c(4700)}=370_{-70-132}^{+70+70}\ \text{MeV}$,
can be safely absorbed into the pole residues $\lambda_{X/Z}$ \cite{meson-loop}.
Thus we take the zero width approximation,
and expect that the predicted masses are reasonable.

On the other hand, the two-point correlation functions $\Pi_{\mu \nu}^{t,a/b}\left(p\right)$
can be calculated at the quark-gluon level via the operator product
expansion method.
We contract the $u$, $d$ and $c$ quark fields
in the correlation functions $\Pi_{\mu \nu}^{t,a/b}\left(p\right)$
with the wick
theorem and obtain the results:
\begin{eqnarray}
\Pi_{\mu \nu}^{t,a}\left(p\right)&=&i\frac{\epsilon^{i j k}\epsilon^{i m n}\epsilon^{i' j' k'}\epsilon^{i' m' n'}}{2}\int d^{4}x e^{ip\cdot x} \nonumber\\
&&\left\{  {\rm Tr}\left[C^{k k'}(x)CU^{j j' T}(x)C\right]{\rm Tr}\left[C^{n' n}(-x)\gamma_{\mu}CD^{m' m T}(-x)C\gamma_{\nu}\right] \right. \nonumber\\
&&+{\rm Tr}\left[C^{k k'}(x)\gamma_{\nu}CU^{j j' T}(x)C\gamma_{\mu}\right]{\rm Tr}\left[C^{n' n}(-x)CD^{m' m T}(-x)C\right] \nonumber\\
&&-t{\rm Tr}\left[C^{k k'}(x)CU^{j j' T}(x)C\gamma_{\mu}\right]{\rm Tr}\left[C^{n' n}(-x)CD^{m' m T}(-x)C\gamma_{\nu}\right] \nonumber\\
&&\left.-t{\rm Tr}\left[C^{k k'}(x)\gamma_{\nu}CU^{j j' T}(x)C\right]{\rm Tr}\left[C^{n' n}(-x)\gamma_{\mu}CD^{m' m T}(-x)C\right]\right\} \ ,
\end{eqnarray}
\begin{eqnarray}
\Pi_{\mu \nu}^{t,b}\left(p\right)&=&-i\frac{\epsilon^{i j k}\epsilon^{i m n}\epsilon^{i' j' k'}\epsilon^{i' m' n'}}{2}\int d^{4}x e^{ip\cdot x} \nonumber\\
&&\left\{  {\rm Tr}\left[C^{k k'}(x)\gamma_{5}CU^{j j' T}(x)C\gamma_{5}\right]{\rm Tr}\left[C^{n' n}(-x)\gamma_{\mu}CD^{m' m T}(-x)C\gamma_{\nu}\right] \right. \nonumber\\
&&+{\rm Tr}\left[C^{k k'}(x)\gamma_{\nu}CU^{j j' T}(x)C\gamma_{\mu}\right]{\rm Tr}\left[C^{n' n}(-x)\gamma_{5}CD^{m' m T}(-x)C\gamma_{5}\right] \nonumber\\
&&-t{\rm Tr}\left[C^{k k'}(x)\gamma_{5}CU^{j j' T}(x)C\gamma_{\mu}\right]{\rm Tr}\left[C^{n' n}(-x)\gamma_{5}CD^{m' m T}(-x)C\gamma_{\nu}\right] \nonumber\\
&&\left.-t{\rm Tr}\left[C^{k k'}(x)\gamma_{\nu}CU^{j j' T}(x)C\gamma_{5}\right]{\rm Tr}\left[C^{n' n}(-x)\gamma_{\mu}CD^{m' m T}(-x)C\gamma_{5}\right]\right\} \ ,
\end{eqnarray}
where the $U_{i j}(x)$, $D_{i j}(x)$ and $C_{i j}(x)$ are the full $u$, $d$ and $c$ quark propagators, respectively,
\begin{eqnarray}
P_{i j}(x)&=&\frac{i\delta_{i j}x\!\!\!/}{2\pi^{2}x^{4}}-\frac{\delta_{i j}\langle\bar{q}q\rangle}{12}-\frac{\delta_{i j}x^{2}\langle\bar{q}g_{s}\sigma Gq\rangle}{192}-\frac{\delta_{i j}x^{2}x\!\!\!/g_{s}^{2}\langle\bar{q}q\rangle^{2}}{7776}-\frac{i g_{s}G_{\alpha\beta}^{n}t_{i j}^{n}(x\!\!\!/\sigma^{\alpha\beta}+\sigma^{\alpha\beta}x\!\!\!/)}{32\pi^{2}x^{2}} \nonumber\\
&&-\frac{\delta_{i j}x^{4}\langle\bar{q}q\rangle\langle GG\rangle}{27648}-\frac{1}{8}\langle\bar{q}_{j}\sigma^{\alpha\beta}q_{i}\rangle\sigma_{\alpha\beta}-\frac{1}{4}\langle\bar{q}_{j}\gamma_{\mu}q_{i}\rangle\gamma^{\mu}+\cdots \ ,\label{propagator-u,d}\\
C_{i j}(x)&=&\frac{i}{(2\pi)^4}\int d^4 ke^{-ik\cdot x}\bigg\{\frac{k\!\!\!/ +m_{c}}{k^{2}-m_{c}^{2}}\delta_{i j}-g_{s}t_{i j}^{n}G_{\alpha\beta}^{n}\frac{(k\!\!\!/+m_{c})\sigma^{\alpha\beta}+\sigma^{\alpha\beta}(k\!\!\!/+m_{c})}{4(k^{2}-m_{c}^{2})^{2}} \nonumber\\
&&+\frac{g_{s}t_{i j}^{n}D_{\alpha}G_{\beta\lambda}^{n}(f^{\lambda\alpha\beta}+f^{\lambda\beta\alpha})}{3(k^{2}-m_{c}^{2})^{4}} \nonumber\\
&&-\frac{g_{s}^{2}(t^{n}t^{m})_{i j}G_{\alpha\beta}^{n}G_{\mu\nu}^{n}(f^{\alpha\beta\mu\nu}+f^{\alpha\mu\beta\nu}+f^{\alpha\mu\nu\beta})}{4(k^{2}-m_{c}^2)^{5}}+\cdots\bigg\} \ ,
\end{eqnarray}
\begin{eqnarray}
f^{\lambda\alpha\beta}&=&(k\!\!\!/+m_{c})\gamma^{\lambda}(k\!\!\!/+m_{c})\gamma^{\alpha}(k\!\!\!/+m_{c})\gamma^{\beta}(k\!\!\!/+m_{c})\ ,\nonumber\\
f^{\alpha\beta\mu\nu}&=&(k\!\!\!/+m_{c})\gamma^{\alpha}(k\!\!\!/+m_{c})\gamma^{\beta}(k\!\!\!/+m_{c})\gamma^{\mu}(k\!\!\!/+m_{c})\gamma^{\nu}(k\!\!\!/+m_{c})\ ,
\end{eqnarray}
the $P_{i j}(x)$ denotes the light quark propagator $U_{i j}(x)$ or $D_{i j}(x)$,
$t^{n}=\frac{\lambda^{n}}{2}$, the $\lambda^{n}$ is the
Gell-Mann matrix,
and $D_{\alpha}=\partial
_\alpha-ig_{s}G_{\alpha}^{n}t^{n}$ \cite{PRT85}.
Then we compute the integrals
both in the coordinate and momentum spaces,
and obtain the correlation functions $\Pi_{\mu \nu}^{t,a/b}\left(p\right)$.
In calculations, we carry out the operator product expansion to the vacuum condensates up to
dimension-10.
The vacuum condensates are the vacuum expectations of the operators $\mathcal{O}_n$, we take the truncations
$n\leq10$ and $k\leq1$ for the operators in a consistent way,
and discard the operators of the orders
$\mathcal{O}( \alpha_s^{k})$ with $k>1$.
In Eq.\eqref{propagator-u,d},
we retain the terms $\langle\bar{q}_{j}\sigma_{\mu\nu}q_{i}\rangle$
and $\langle\bar{q}_{j}\gamma_{\mu}q_{i}\rangle$
originating from the Fierz re-arrangement of the $\langle q_{i}\bar{q}_{j}\rangle$
to absorb the gluons emitted from the heavy quark lines
so as to extract the mixed
condensates and four-quark condensates
$\langle\bar{q}g_{s}\sigma Gq\rangle$ and $g_{s}^{2}\langle\bar{q}q\rangle^{2}$, respectively.
One can consult Ref.\cite{axialvector}
for some technical details in the operator product expansion.
Once the analytical expressions of the correlation functions $\Pi_{\mu \nu}^{t,a/b}\left(p\right)$
are gotten, we can obtain the corresponding correlation functions $\Pi_{0}^{t,a/b}\left(p\right)$
of the scalar and pseudoscalar tetraquark states with:
$\Pi_{0}^{t,a/b}\left(p\right)=\frac{p^{\mu}\Pi_{\mu \nu}^{t,a/b}\left(p\right)}{p^2 p_\nu}$.
The QCD spectral densities $\rho^{Z_{0}^{\,t,a/b}}\left(s\right)$
of the scalar and pseudoscalar tetraquark states are obtained successfully
through dispersion relation.

After getting the explicit expressions of the QCD spectral densities $\rho^{Z_{0}^{\,t,a/b}}\left(s\right)$,
we take the quark-hadron duality bellow the continuum threshold value $s_0$ and perform Borel transform with
respect to the variable $P^2 =-p^2$ to obtain the following QCD sum rules:
\begin{eqnarray}\label{PoleResidue}
\lambda_{Z_{0}^{\,t,a/b}}^{2}\exp\left(-\frac{M_{Z_{0}^{\,t,a/b}}^{2}}{T^{2}}\right)
&=&\int_{4m_{c}^{2}}^{s_{0}}ds\rho^{Z_{0}^{\,t,a/b}}\left(s\right)\exp\left(-\frac{s}{T^{2}}\right)\ ,
\end{eqnarray}
where
\begin{eqnarray}
\rho^{Z_{0}^{\,t,a/b}}\left(s\right)&=& \rho_{0}^{\,t,a/b}\left(s\right)+\rho_{3}^{\,t,a/b}\left(s\right)+\rho_{4}^{\,t,a/b}\left(s\right)
+\rho_{5}^{\,t,a/b}\left(s\right)+\rho_{6}^{\,t,a/b}\left(s\right)+\rho_{7}^{\,t,a/b}\left(s\right) \nonumber\\
&&+\rho_{8}^{\,t,a/b}\left(s\right)+\rho_{10}^{\,t,a/b}\left(s\right)\ ,
\end{eqnarray}
the subscripts 0, 3, 4, 5, 6, 7, 8 and 10 denote the dimensions of the vacuum condensates
in the operator product expansion,
the $T^2$ denotes the Borel parameter.
We collect the spectral densities $\rho^{Z_{0}^{\,t,a/b}}\left(s\right)$ explicitly
in the appendix.

Differentiate Eq.\eqref{PoleResidue} with respect to $\frac{1}{T^2}$
and eliminate the pole residues $\lambda_{Z_{0}^{\,t,a/b}}$, we obtain the
QCD sum rules for the masses of the scalar and pseudoscalar tetraquark states,
\begin{eqnarray}\label{mass}
M_{Z_{0}^{\,t,a/b}}^{2}&=&\frac{\int_{4m_{c}^{2}}^{s_{0}}ds\frac{d}
{d\left(-1/T^{2}\right)}\rho^{Z_{0}^{\,t,a/b}}\left(s\right)\exp\left(-\frac{s}{T^{2}}\right)} {\int_{4m_{c}^{2}}^{s_{0}}ds\rho^{Z_{0}^{\,t,a/b}}\left(s\right)\exp\left(-\frac{s}{T^{2}}\right)}\ .
\end{eqnarray}

\section{Numerical results and discussions}
In this section, we perform the numerical analysis.
For the hadron mass,
it is independent of the energy scale because of its observability.
However, in our calculations,
we discard the perturbative corrections and the operators of the
orders $\mathcal{O}( \alpha_s^{k})$ with $k>1$
or the dimensions $n>10$,
and factorize the higher dimension operators into non-factorizable low
dimension operators with the same quantum numbers of the vacuum.
In addition,
the variation of the heavy mass $m_c$ depending on the energy scale
leads to change of integral range
$4m_c^2-s_0$ of the variable $ds$.
So we have to consider the energy scale dependence of the QCD sum rules.
The input parameters at the QCD side are taken to be the standard condensate values
$\langle\bar{q}q\rangle=-(0.24\pm0.01\,\text{GeV})^{3}$,
$\langle\bar{q}g_{s}\sigma Gq\rangle=m_{0}^{2}\langle\bar{q}q\rangle$,
$m_{0}^{2}=(0.8\pm0.1)\,\text{GeV}^{2}$,
$\langle\frac{\alpha_{s}GG}{\pi}\rangle=(0.33\,\text{GeV})^{4}$
at the energy scale  $\mu=1\,\text{GeV}$ from the Gell-Mann-Oakes-Renner
relation \cite{PRT85,NPB147,175}, and the $\overline{MS}$ mass $m_{c}(m_c)=(1.275\pm0.025)\,\rm{GeV}$
from the Particle Data Group \cite{XYZ}.
Moreover, we neglect the light quark masses and take into account
the energy-scale dependence of the quark condensate, mixed quark condensate
and $\overline{MS}$ mass from the renormalization group equation,
\begin{eqnarray}
\langle\bar{q}q\rangle(\mu)&=&\langle\bar{q}q\rangle(Q)[\frac{\alpha_{s}(Q)}{\alpha_{s}(\mu)}]^{\frac{4}{9}}\,,\nonumber\\
\langle\bar{q}g_{s}\sigma Gq\rangle(\mu)&=&\langle\bar{q}g_{s}\sigma
Gq\rangle(Q)[\frac{\alpha_{s}(Q)}{\alpha_{s}(\mu)}]^{\frac{2}{27}}\,,\nonumber\\
m_c(\mu)&=&m_c(m_c)\left[\frac{\alpha_{s}(\mu)}{\alpha_{s}(m_c)}\right]^{\frac{12}{25}} \, ,\nonumber\\
\alpha_s(\mu)&=&\frac{1}{b_0t'}\left[1-\frac{b_1}{b_0^2}\frac{\log t'}{t'} +\frac{b_1^2(\log^2{t'}-\log{t'}-1)+b_0b_2}{b_0^4t'^2}\right]\, ,
\end{eqnarray}
where $t'=\log \frac{\mu^2}{\Lambda^2}$, $b_0=\frac{33-2n_f}{12\pi}$, $b_1=\frac{153-19n_f}{24\pi^2}$, $b_2=\frac{2857-\frac{5033}{9}n_f+\frac{325}{27}n_f^2}{128\pi^3}$,  $\Lambda=213\,\rm{MeV}$, $296\,\rm{MeV}$
and  $339\,\rm{MeV}$ for the flavors  $n_f=5$, $4$ and $3$, respectively \cite{XYZ}.

In Eq.\eqref{mass}, there are two free parameters: the Borel Parameter $T^2$ and the continuum threshold value $s_0$.
The extracted hadron mass is a function of the Borel parameter $T^2$ and the continuum threshold value $s_0$.
To obtain a reliable mass sum rule analysis,
we impose two criteria on the hidden-charm tetraquark states
to choose suitable working ranges for these two free parameters.
The first criterion is the pole dominance on the phenomenological side,
which require the pole contributions to be about $(40-60)\%$.
The pole contribution (PC) is defined as,
\begin{eqnarray}
\text{PC}&=&\frac{\int_{4m_{c}^{2}}^{s_{0}}ds\rho^{Z_{0}^{\,t,a/b}}\left(s\right)\exp\left(-\frac{s}{T^{2}}\right)} {\int_{4m_{c}^{2}}^{\infty}ds\rho^{Z_{0}^{\,t,a/b}}\left(s\right)\exp\left(-\frac{s}{T^{2}}\right)}\ .
\end{eqnarray}
The second criterion is the convergence of the operator product expansion. To judge the
convergence, we calculate the contributions $D_i$ in the
operator product expansion with the formula,
\begin{eqnarray}
D_i&=&\frac{\int_{4m_{c}^{2}}^{s_{0}}ds\rho_{i}^{Z_{0}^{\,t,a/b}}(s)\exp\left(-\frac{s}{T^{2}}\right)}
{\int_{4m_{c}^{2}}^{s_{0}}ds\rho^{Z_{0}^{\,t,a/b}}\left(s\right)\exp\left(-\frac{s}{T^{2}}\right)}\ ,
\end{eqnarray}
where the index $i$ denotes the dimension of the vacuum condensates.

To search for the continuum threshold value $s_0$ more precisely,
we take into account the mass gaps
between the ground states and the first radial excited states,
which are usually taken as $(0.4-0.6)\,\text{GeV}$ in the tetraquark sector.
For examples, the $Z(4430)$ is tentatively
assigned to be the first radial excitation of the $Z_{c}(3900)$
according to the analogous decays, $Z_{c}(3900)^{\pm} \longrightarrow  J/\psi \pi^{\pm}$, $Z(4430)^{\pm} \longrightarrow \psi'\pi^{\pm}$
and the mass differences $M_{Z(4430)}-M_{Z_{c}(3900)}=576\,\text{MeV}$, $M_{\psi'}-M_{J/\psi}=589\,\text{MeV}$ [41-44];
the $X(3915)$ and $X(4500)$ are assigned  to be the ground state and the first radial excited state of the axialvector-diquark-axialvector-antidiquark type scalar $cs\bar{c}\bar{s}$ tetraquark states, respectively, and their mass difference is $M_{X(4500)}-M_{X(3915)}=588\,\text{MeV}$ \cite{EPJC77}. The relation
\begin{eqnarray}\label{s0-relation}
\sqrt{s_{0}}&=&M_{X/Y/Z}+(0.4-0.6)\,\text{GeV}\ ,
\end{eqnarray}
serves as a constraint on the masses of the hidden-charm tetraquark states.

\begin{figure}[htp]
\centering
\includegraphics[totalheight=5cm,width=7cm]{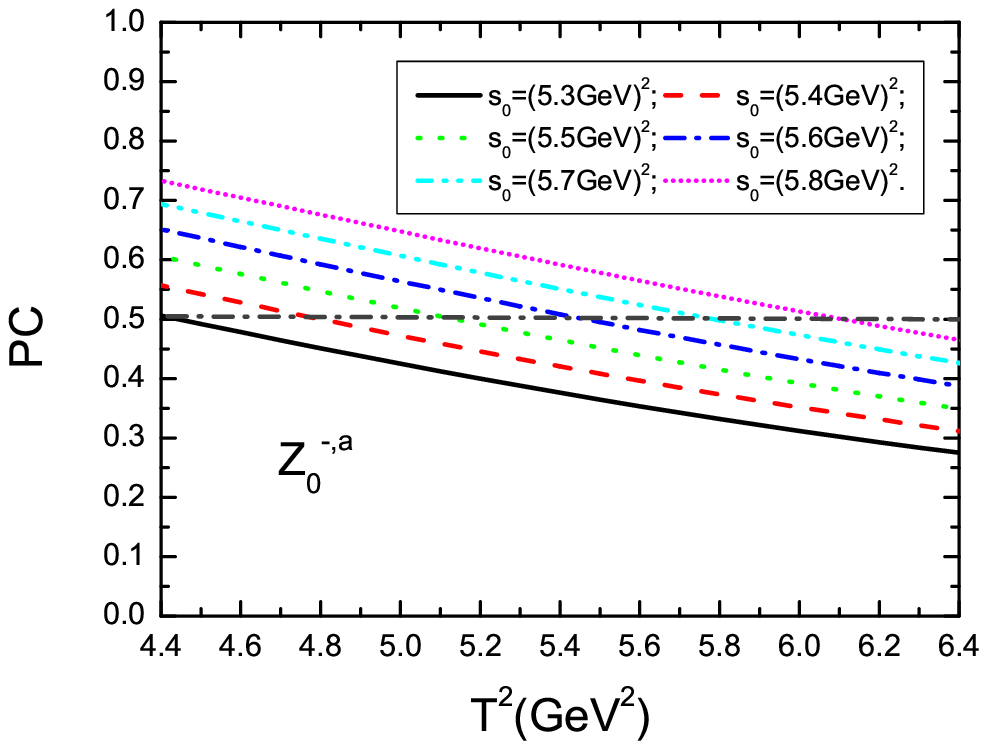}
\includegraphics[totalheight=5cm,width=7cm]{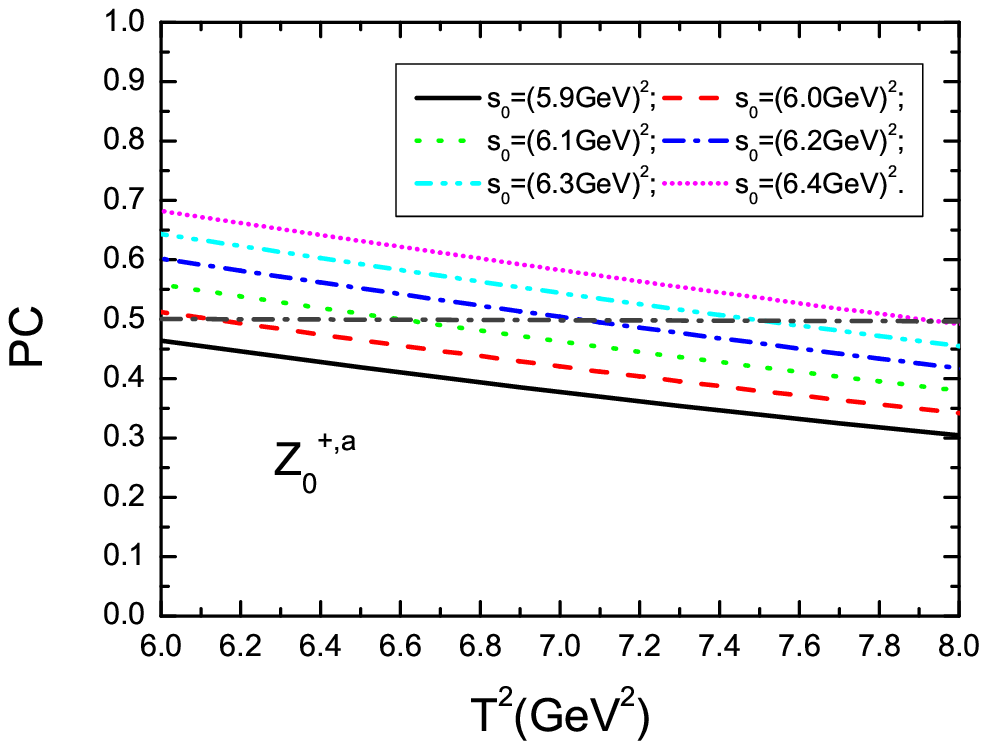}
\includegraphics[totalheight=5cm,width=7cm]{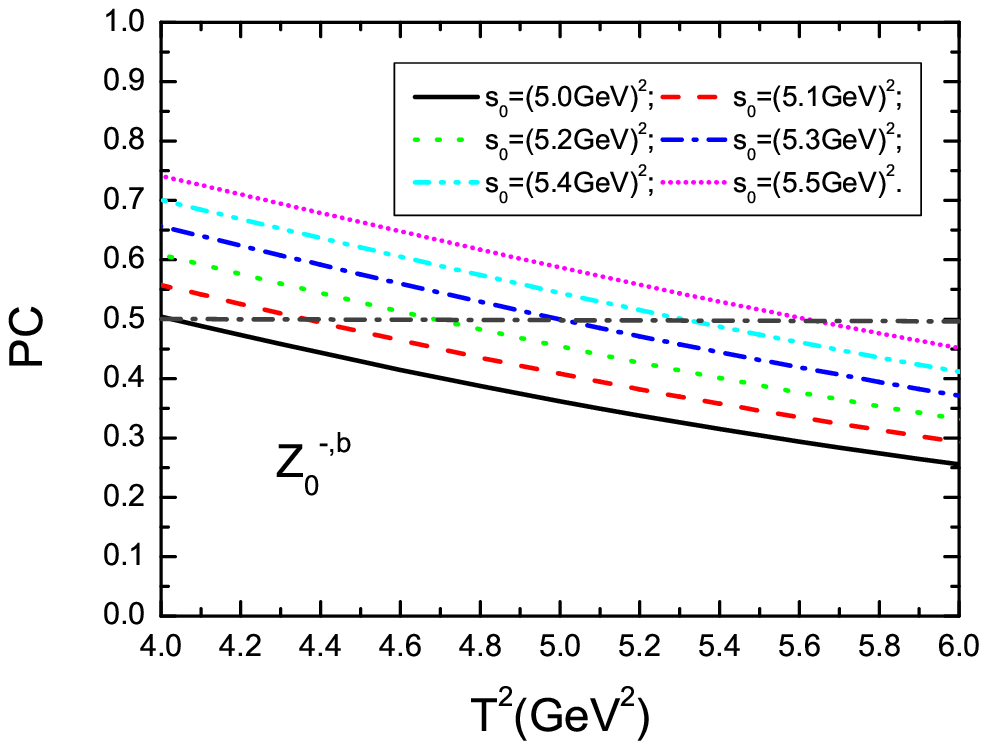}
\includegraphics[totalheight=5cm,width=7cm]{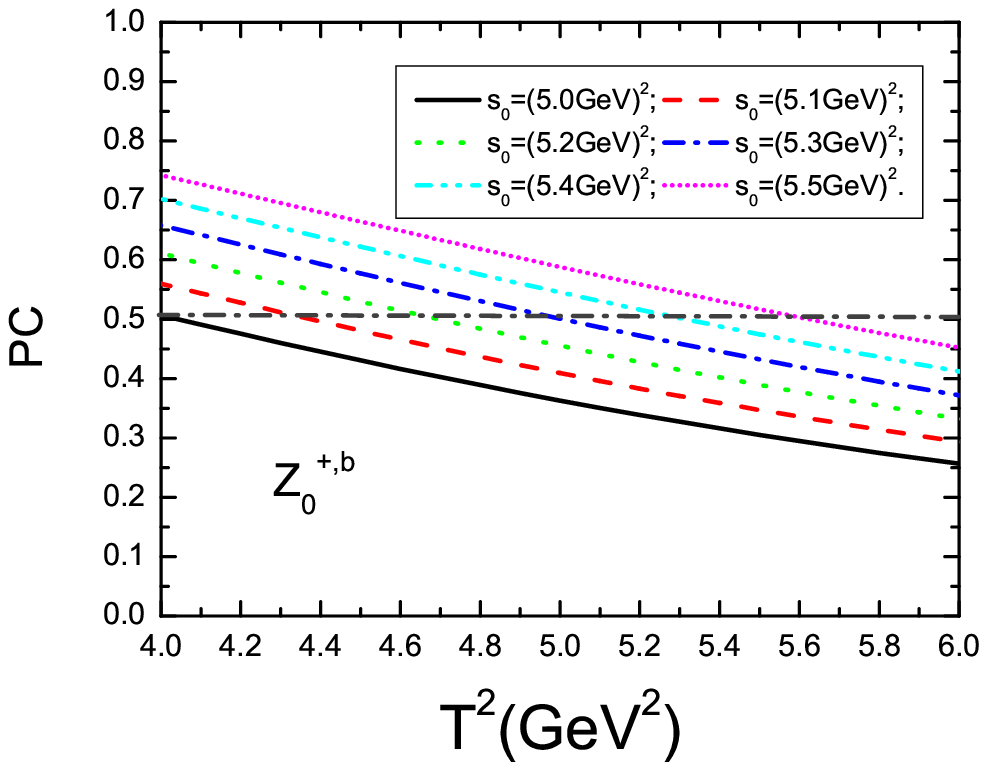}
\caption{The pole contributions with variations of the Borel parameters $T^{2}$ and threshold parameters $s_{0}$.}\label{fig:fig1}
\end{figure}

In this article, we take the energy scale $\mu$ as a free parameter and
evolve all the input parameters in the QCD spectral density to
the special energy scale determined by the empirical formula,
\begin{eqnarray}\label{Escalar}
\mu&=&\sqrt{M_{X/Y/Z}^{2}-\left(2{\mathbb{M}}_{c}\right)^{2}}\ ,
\end{eqnarray}
with the effective c-quark mass ${\mathbb{M}}_{c}=1.82\,\text{GeV}$.
The heavy tetraquark system could be described by a double-well potential with two
light quarks $q'\bar{q}$ lying in the two wells respectively.
In the heavy quark limit, the $c$ quark can be taken as a static well potential,
which binds the light quark $q'$ to form a diquark in the color antitriplet channel
or binds the light antiquark $\bar{q}$ to form a meson in the color singlet channel
(or a meson-like state in the color octet channel).
Then the heavy tetraquark states
are characterized by the effective heavy quark masses ${\mathbb{M}}_{c}$
(or constituent quark masses) and the
virtuality $V=\sqrt{M_{X/Y/Z}^{2}-\left(2{\mathbb{M}}_{c}\right)^{2}}$.
It is natural to take the energy
scale $\mu=V$.
For a better understanding of the energy scale dependence in Eq.\eqref{Escalar},
one can refer to Refs.[35,36,46-49], where the authors study the energy scale dependence of the
QCD sum rules for the hidden-charm tetraquark states and molecular states in detail,
and suggest the above energy scale formula for the first time.
The energy scale formula works well for the $X(3872)$, $Z_c(3885/3900)$, $X^*(3860)$,
$Y(3915)$, $Z_c(4020/4025)$,   $Z(4430)$, $X(4500)$, $Y(4630/4660)$, $X(4700)$
in the scenario of tetraquark states.
Actually, the formula put another constraint on the masses of the hidden-charm tetraquark states.
In our calculations, we observe that the values  of the
masses $M_Z$ decrease slightly with  increase of the energy scales   $\mu$
from QCD sum rules in Eq.\eqref{mass},
while Eq.\eqref{Escalar} indicates
that the value of the masses  $M_Z$ increase when the energy scales
$\mu$ increase. Thus there exist optimal energy scales, which lead to  reasonable  masses $M_{Z}$.

\begin{figure}[htp]
\centering
\includegraphics[totalheight=5cm,width=7cm]{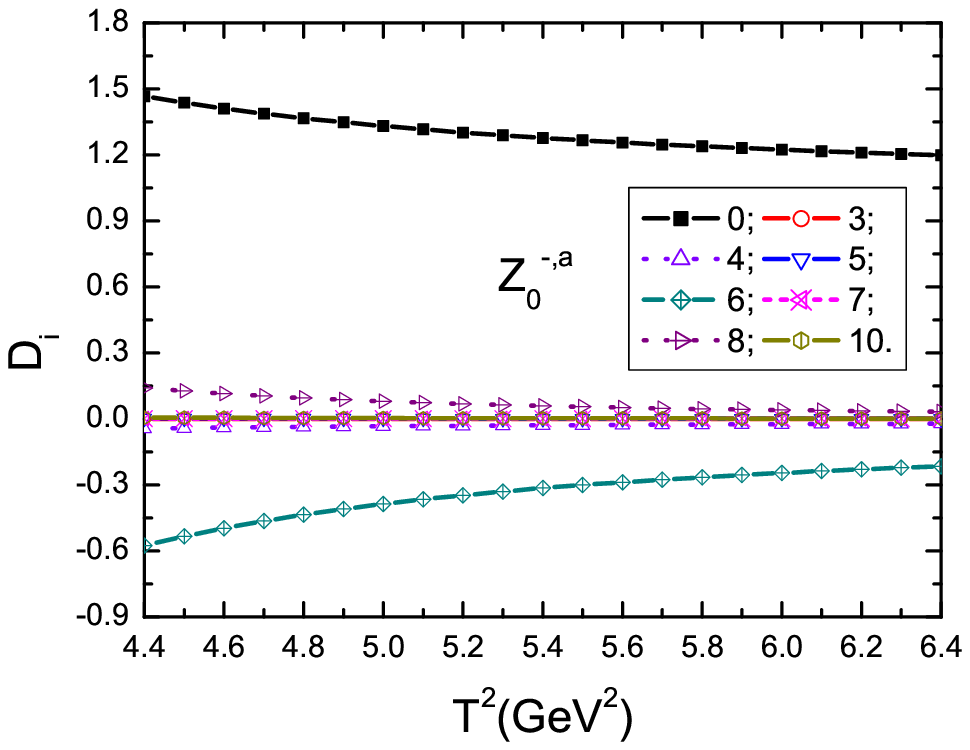}
\includegraphics[totalheight=5cm,width=7cm]{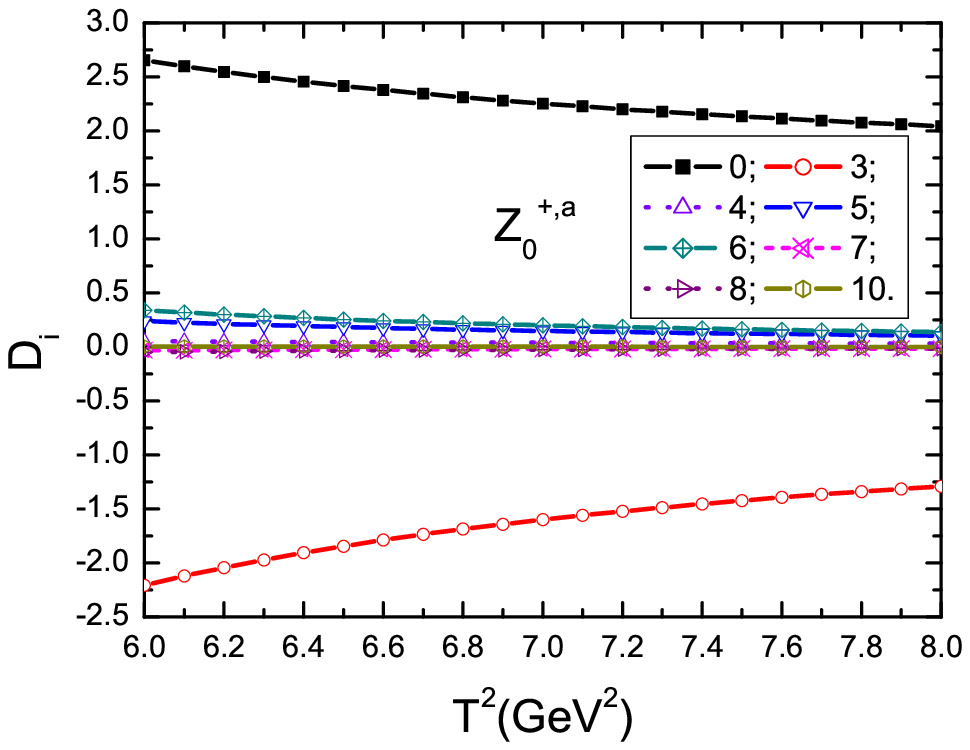}
\includegraphics[totalheight=5cm,width=7cm]{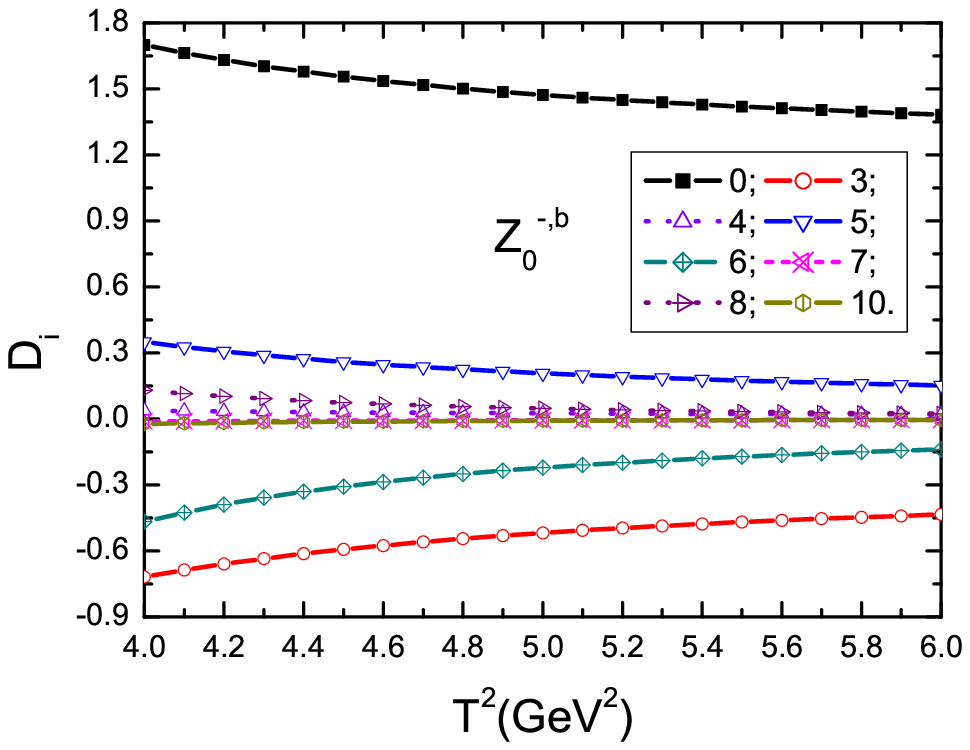}
\includegraphics[totalheight=5cm,width=7cm]{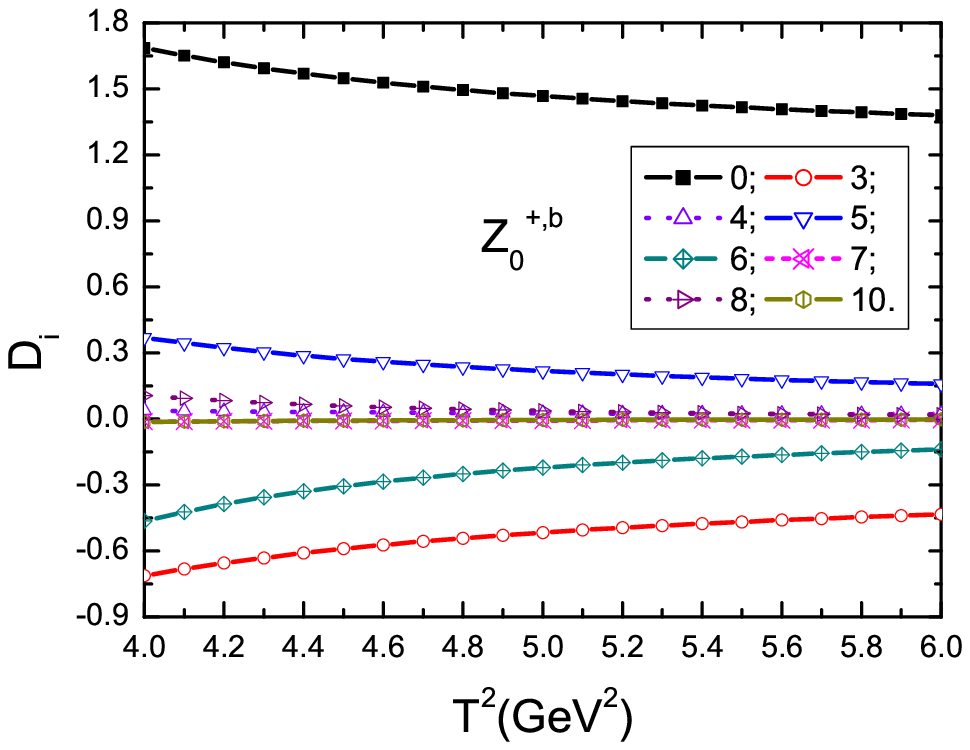}
\caption{The contributions of different terms in the operator product expansion with variations of the Borel parameters $T^{2}$, where the $0$, $3$, $4$, $5$, $6$, $7$, $8$, $10$ denote the dimensions of the vacuum condensates.}\label{fig:fig2}
\end{figure}

\begin{figure}[htp]
\centering
\includegraphics[totalheight=5cm,width=7cm]{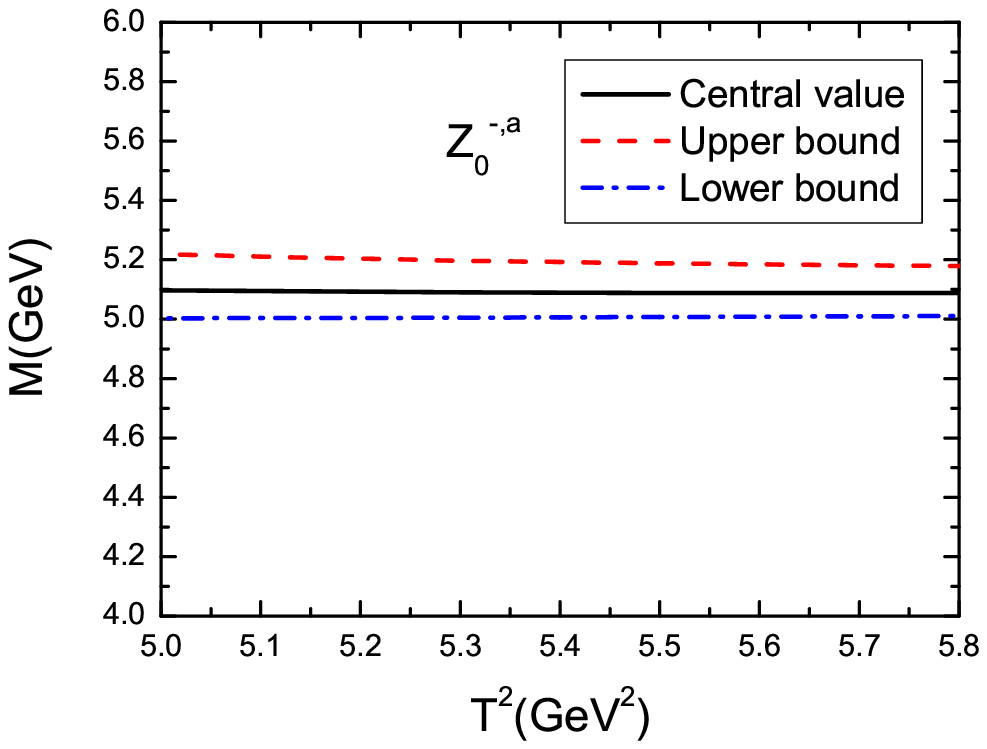}
\includegraphics[totalheight=5cm,width=7cm]{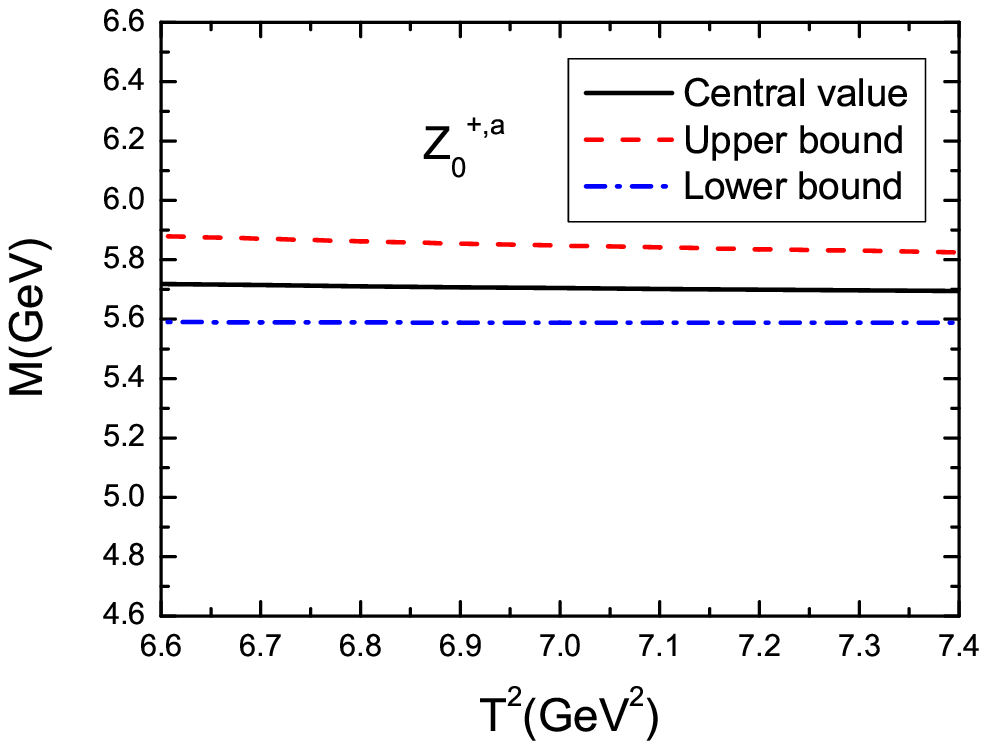}
\includegraphics[totalheight=5cm,width=7cm]{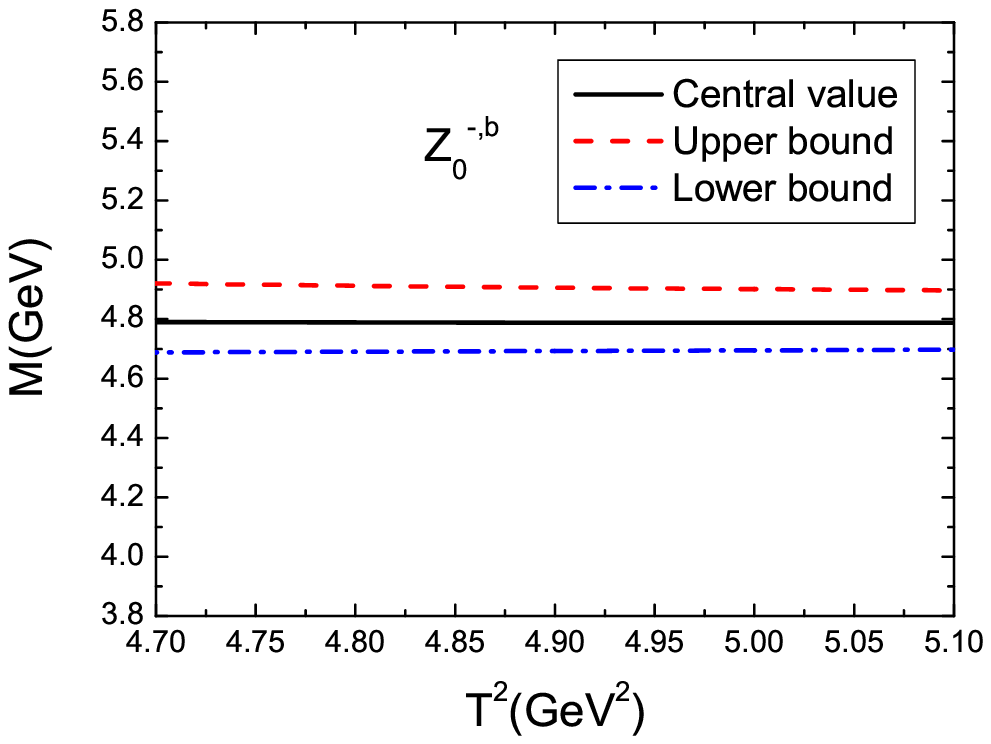}
\includegraphics[totalheight=5cm,width=7cm]{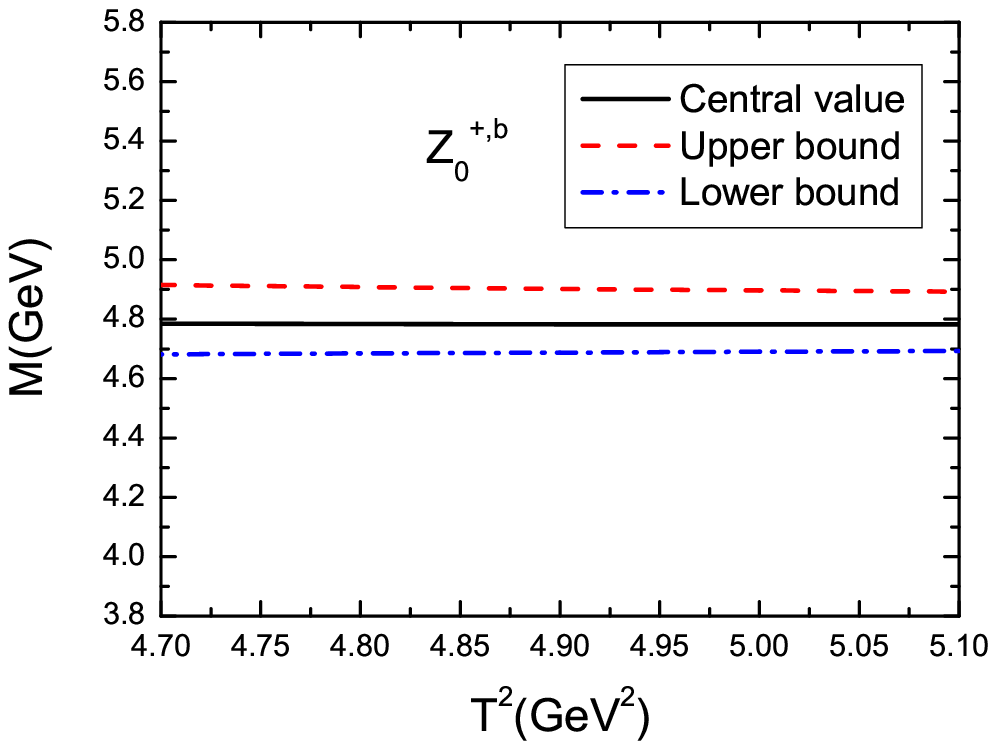}
\caption{The masses with variations of the Borel parameters $T^{2}$.}\label{fig:fig3}
\end{figure}

\begin{figure}[htp]
\centering
\includegraphics[totalheight=5cm,width=7cm]{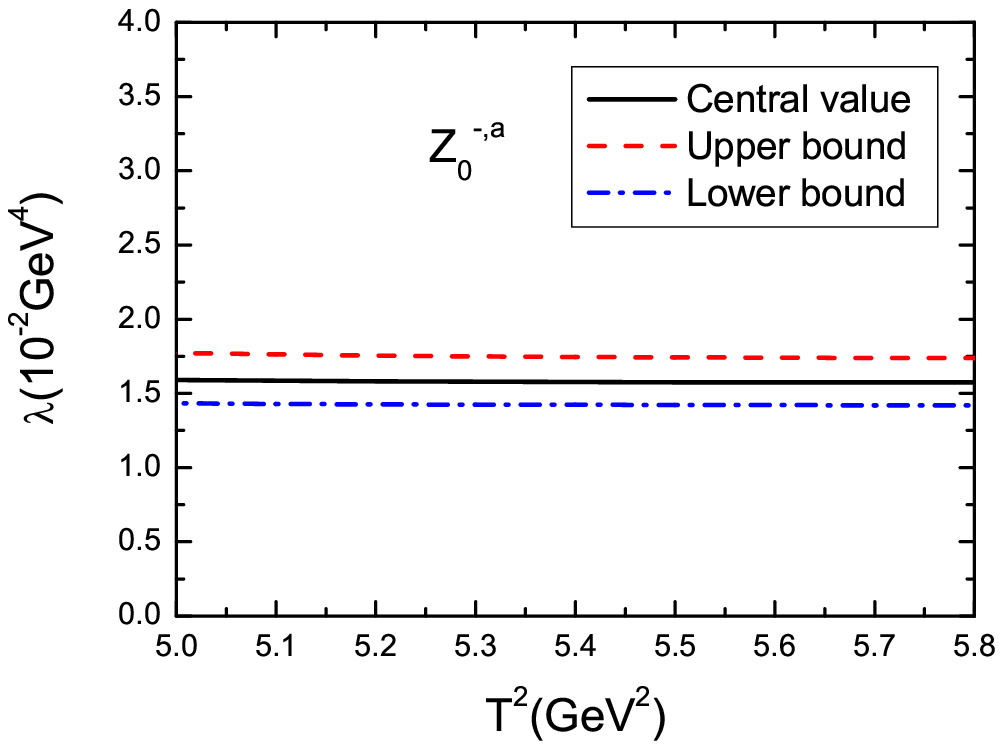}
\includegraphics[totalheight=5cm,width=7cm]{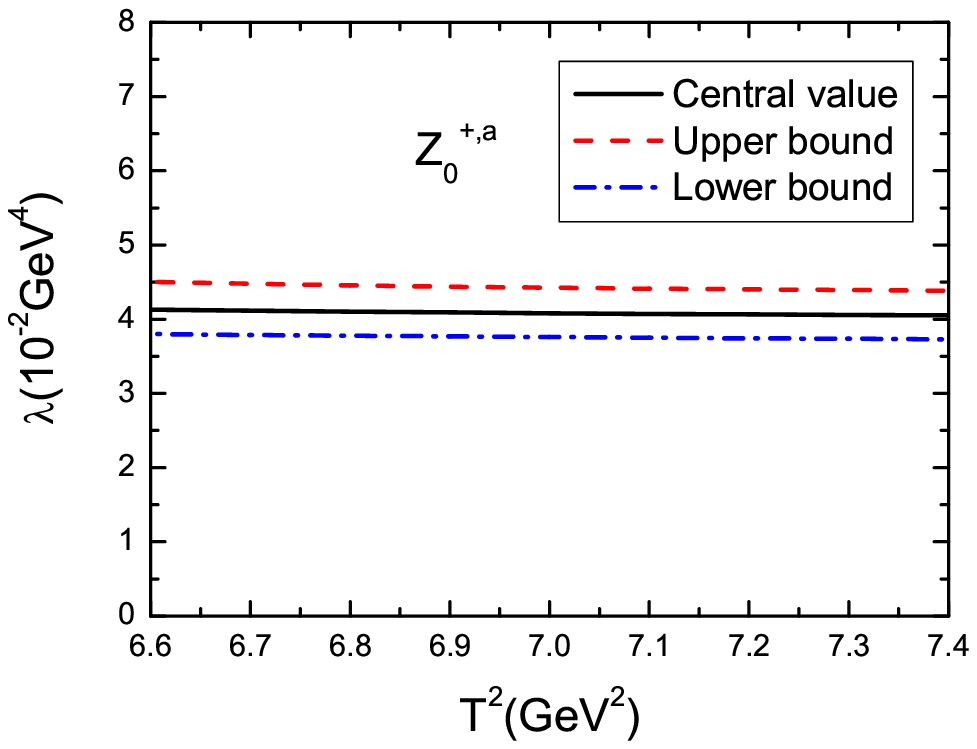}
\includegraphics[totalheight=5cm,width=7cm]{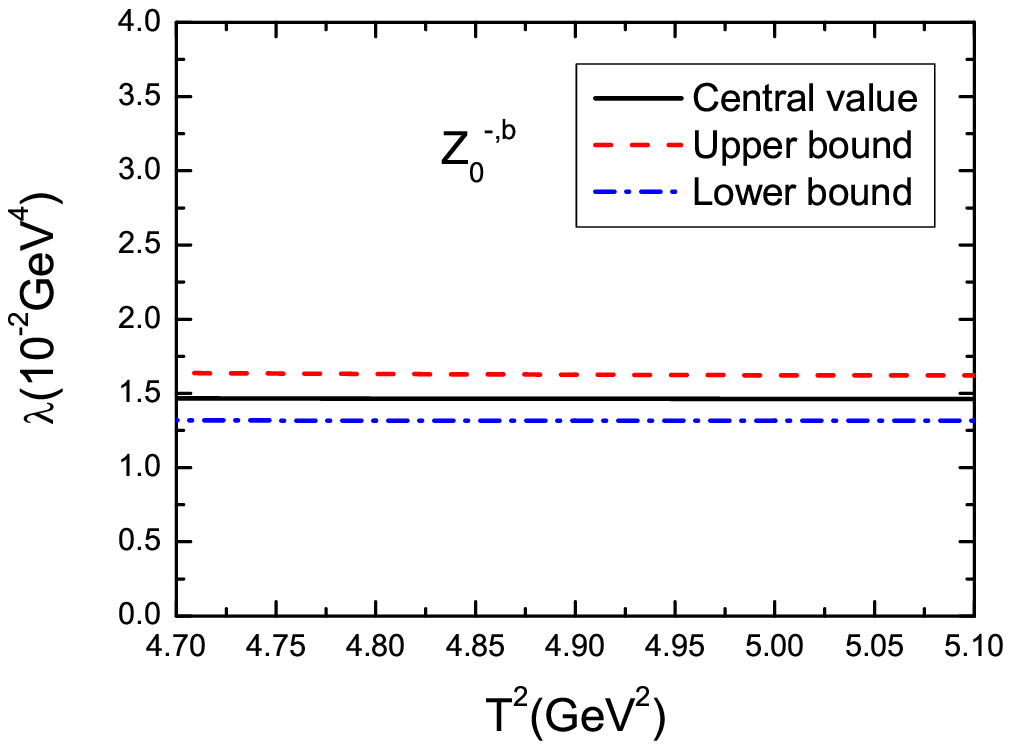}
\includegraphics[totalheight=5cm,width=7cm]{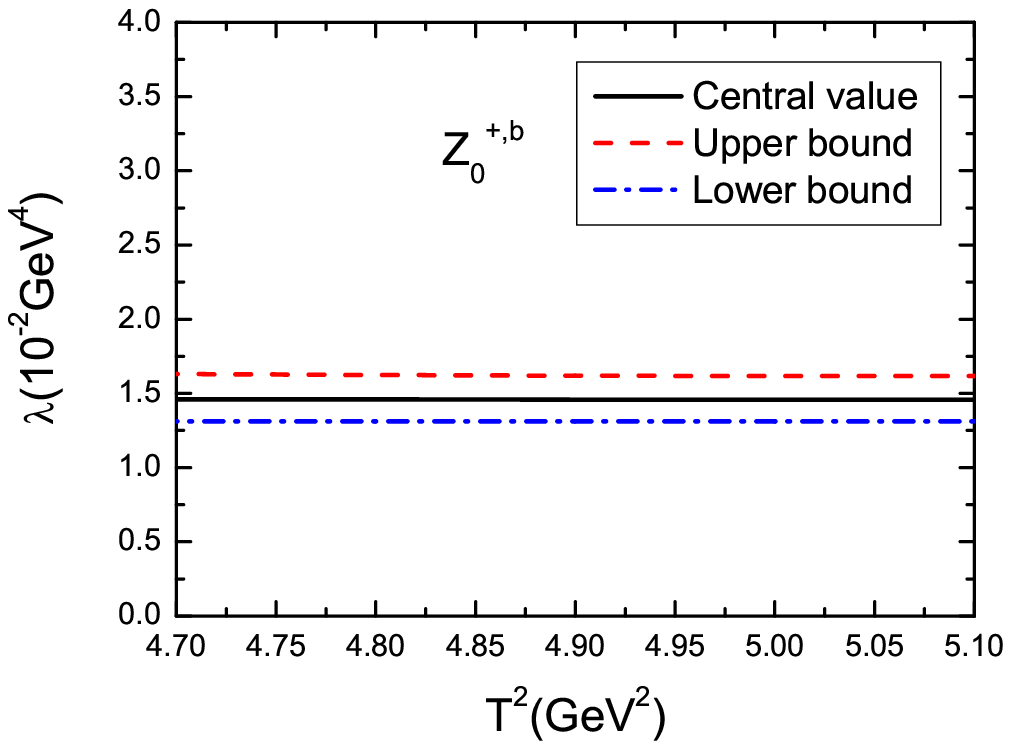}
\caption{The pole residues with variations of the Borel parameters $T^{2}$.}\label{fig:fig4}
\end{figure}

In Fig.\ref{fig:fig1}, we show the variations of the pole contributions with respect to the Borel parameters $T^2$ for different values of the continuum thresholds $s_0$ at the energy scales $\mu=3.6\, \text{GeV}$, $4.4\, \text{GeV}$, $3.2\, \text{GeV}$ and $3.2\, \text{GeV}$ for the tetraquark states $ Z_{0}^{-\,,a}$, $ Z_{0}^{+\,,a}$, $ Z_{0}^{-\,,b}$ and $ Z_{0}^{+\,,b}$, respectively.
In Fig.\ref{fig:fig2}, the contributions of
different terms in the operator product expansion are plotted with variations of the Borel parameters $T^2$
at the parameters $s_{0}=5.6\,\text{GeV}$, $\mu=3.6\,\text{GeV}$; $s_{0}=6.2\,\text{GeV}$, $\mu=4.4\,\text{GeV}$;
$s_{0}=5.3\,\text{GeV}$, $\mu=3.2\,\text{GeV}$ and $s_{0}=5.3\,\text{GeV}$, $\mu=3.2\,\text{GeV}$ for the tetraquark
states $ Z_{0}^{-\,,a}$, $ Z_{0}^{+\,,a}$, $ Z_{0}^{-\,,b}$ and $ Z_{0}^{+\,,b}$, respectively.
From the figures,
we can choose the optimal Borel parameters and threshold parameters to satisfy the two criteria of the QCD sum rules.
To explain the procedure, we take the scalar tetraquark state $ Z_{0}^{-\,,a}$ with $J^{PC}=0^{++}$ as an example.
From the first panel of Fig.\ref{fig:fig1}, we can see that the values $\sqrt{s_{0}}\leq5.4\,\text{GeV}$
are too small to satisfy the pole dominance condition and
result in reasonable Borel windows.
In the first panel of Fig.\ref{fig:fig2},
the contributions of different terms change quickly
with respect to the Borel parameter at the region $T^2<5.0\,\text{GeV}^{2}$,
which does not warrant platform for the mass.
At the value $T^2=5.0\,\text{GeV}^{2}$, the $D_0$, $D_3$, $D_4$, $D_5$, $D_6$, $D_7$, $D_8$, $D_{10}$
are 1.331, 0.000, -0.034, 0.003, -0.386, 0.002, 0.080, 0.004 respectively for the tetraquark state $ Z_{0}^{-\,,a}$
and the total contributions are normalized to be 1.
Accordingly, the $T^{2}\geq5.0\,\text{GeV}^{2}$ is taken tentatively,
the perturbative term plays an important role,
and the convergent behavior in the operator product expansion is very good.
If we take the values $\sqrt{s_{0}}=(5.5-5.7)\,\text{GeV}$ and $T^{2}=(5.0-5.8)\,\text{GeV}^{2}$,
the pole contribution is about $(42-61)\%$ for the state $ Z_{0}^{-\,,a}$.
The pole dominance condition is well satisfied.
Similarly, we obtain the
Borel parameters, continuum thresholds and the pole
contributions for all tetraquark states $ Z_{0}^{\pm\,,a/b}$, which are shown explicitly in Table \ref{tab:tablenotes}.

\begin{table}[!ht]
\begin{center}
\begin{tabular}{|c|c|c|c|c|c|c|}
  \hline
   \hline
  & $\mu(\text{GeV})$ & $T^{2}(\text{GeV}^{2})$ & $\sqrt{s_{0}}(\text{GeV})$  & pole & $M_{Z}(\text{GeV})$ &  $\lambda_{Z}(\text{GeV}^{4})$\\
   \hline
   $Z_{0}^{-\,,a}\left(0^{++}\right)$ &$3.6$& $5.0-5.8$ & $5.6\pm0.1$ & $\left(42-61\right)\%$ & $5.09_{-0.08}^{+0.13}$ & $\left(1.58_{-0.16}^{+0.20}\right)\times10^{-2}$\\
   \hline
   $Z_{0}^{+\,,a}\left(0^{+-}\right)$ &$4.4$& $6.6-7.4$ & $6.2\pm0.1$ & $\left(43-58\right)\%$ & $5.70_{-0.12}^{+0.18}$ & $\left(4.08_{-0.35}^{+0.42}\right)\times10^{-2}$\\
   \hline
   $Z_{0}^{-\,,b}\left(0^{-+}\right)$ &$3.2$& $4.7-5.1$ & $5.3\pm0.1$ & $\left(44-59\right)\%$ & $4.79_{-0.09}^{+0.13}$ & $\left(1.45_{-0.15}^{+0.17}\right)\times10^{-2}$\\
  \hline
  $Z_{0}^{+\,,b}\left(0^{--}\right)$ &$3.2$& $4.7-5.1$ & $5.3\pm0.1$ & $\left(44-59\right)\%$ & $4.78_{-0.09}^{+0.13}$ & $\left(1.44_{-0.15}^{+0.17}\right)\times10^{-2}$\\
   \hline
  \hline
\end{tabular}
\end{center}
\caption{The energy scales, Borel parameters, continuum threshold
parameters, pole contributions, masses and pole
residues for the scalar and pseudoscalar tetraquark states.}\label{tab:tablenotes}
\end{table}

We take into account all uncertainties of the input parameters, and obtain the values of the
masses and pole residues of the tetraquark states, which are shown in Table \ref{tab:tablenotes}
and Figs.\ref{fig:fig3}-\ref{fig:fig4}.
From Figs.\ref{fig:fig3}-\ref{fig:fig4}, we can see that the Borel platforms exist.
On the other hand,
from Table \ref{tab:tablenotes},
we can see that
the energy scale formula $ \mu =\sqrt{M_{X/Y/Z}^2-({2\mathbb{M}}_c)^2}$
and the relation $\sqrt{s_{0}}=M_{X/Y/Z}+(0.4-0.6)\,\text{GeV}$ are well satisfied.
The numerical results indicate that none of
the tetraquark states $ Z_{0}^{\pm\,,a/b}$ is the lowest hidden charmed tetraquark state,
whose mass is about $3.82\,\text{GeV}$ \cite{lowest-tetraquark};
the charge conjugation
partners have almost degenerate masses for the pseudoscalar tetraquark states $ Z_{0}^{\pm\,,b}$,
while there is a considerably large energy gap about $610\,\text{MeV}$  between the masses of the $C=+$
and $C=-1$ scalar tetraquark states $ Z_{0}^{\pm\,,a}$;
the mass predictions of the scalar tetraquark states are larger than
the counterparts of the pseudoscalar tetraquark states
because the scalar  tetraquark states
have the $C\otimes C$ type substructure and
the pseudoscalar  tetraquark states have more stable $C\gamma_5 \otimes C$ type and
$C\otimes \gamma_5 C$ type substructures,
which can be found in Eqs.\eqref{scalar}-\eqref{pseudoscalar}.
A meson may have a lot of Fock stastes with different constituents,
such as $\bar{q}q$, $\bar{q}qg$, $\bar{q}q\bar{q}q$, etc.
These mesons we study in this article have non-vanishing couplings
with the tetraquark currents, thus these mesons contain the tetraquark constituent.
The present predictions can be confronted with the
experimental data in the future at the BESIII, LHCb and Belle-II.

Using the masses obtained above, we study the possible  hadronic decay patterns of
the scalar and pseudoscalar hidden-charm tetraquark states $Z_{0}^{\pm\,,a/b}$.
It's known that a hidden-charm tetraquark state composed of a diquark and antidiquark pair can decay easily into a pair of
open-charm $D$ mesons or one charmonium state plus a light meson through quark rearrangement.
Such two-body strong decays are Okubo-Zweig-Iizuka super-allowed.
Considering the conservation of the angular momentum, parity, charge conjugation and isospin, we list out the possible strong
decays of the $Z_{0}^{\pm\,,a/b}$,
\begin{eqnarray}
Z_{0}^{-\,,a}\left(0^{++}\right)&\longrightarrow& \eta_c\pi^+,\, J/\psi\rho^+(770),\, \psi(3770)\rho^+(770),\, \chi_{c0}a_0^{+}(980),\,  \chi_{c1}a_1^+(1260),\, h_c b_1 ^{+}(1235),\, \nonumber\\
    &&\bar{D}^0D^+,\, \bar{D}^{*0}(2007)D^{*+}(2010),\, \bar{D}_0^{*0}(2400)D_0^{*+}(2400),\, \bar{D}_1^0(2420)D_1^+(2420),\nonumber\\
    &&\bar{D}_1^0(2430)D_1^+(2430),\nonumber\\
Z_{0}^{+\,,a}\left(0^{+-}\right)&\longrightarrow& J/\psi\pi_{1}^+(1400),\, \psi(3770)\pi_{1}^+(1400),\, \chi_{c1}b_{1}^+(1235),\, h_c a_1^+(1260),\, \bar{D}^0D^+,\,  \nonumber\\
&&\bar{D}^{*0}(2007)D^{*+}(2010),\, \bar{D}_0^{*0}(2400)D_0^{*+}(2400),\, \bar{D}_1^0(2420)D_1^+(2420),\, \nonumber\\
&&\bar{D}_1^0(2430)D_1^+(2430),\nonumber\\
Z_{0}^{-\,,b}\left(0^{-+}\right)&\longrightarrow& \eta_ca_0^{+}(980),\,  J/\psi b_{1}^+(1235),\, \chi_{c0}\pi^+,\, h_c\rho^+(770),\, \bar{D}_0^{*0}(2400)D^+,\, \bar{D}^0D_0^{*+}(2400),\,\nonumber\\
&&\bar{D}_1^0(2420)D^{*+}(2010),\, \bar{D}_1^0(2430)D^{*+}(2010),\nonumber\\
Z_{0}^{+\,,b}\left(0^{--}\right)&\longrightarrow& J/\psi a_1^+(1260),\, \chi_{c1}\rho^+(770),\, \bar{D}_0^{*0}(2400)D^+,\, \bar{D}^0D_0^{*+}(2400),\nonumber\\ &&\bar{D}_1^0(2420)D^{*+}(2010),\, \bar{D}_1^0(2430)D^{*+}(2010).
\end{eqnarray}
Under the restriction of charge conjugation, the decay modes of double open-charm $D$ mesons
are dominant for the scalar and pseudoscalar tetraquark states with negative charge conjugation.
For the scalar tetraquark states, the $ Z_{0}^{-\,,a}$ is much narrower than the $ Z_{0}^{+\,,a}$,
as the mass of the $ Z_{0}^{-\,,a}$ is much smaller.
Thus, compared to the $ Z_{0}^{+\,,a}$, the $ Z_{0}^{-\,,a}$ will be prime candidate for observation.

\section{Conclusion}
In this article, based on the diquark configuration, we use the scalar, pseudoscalar, axialvector
diquarks and their corresponding antidiquarks to construct the vector and axial-vector
interpolating tetraquark currents,
which can couple to the scalar and
pseudoscalar tetraquark states respectively.
Then we distinguish the charge conjugations of the interpolating
currents.
In calculations, we consider the contributions of the vacuum condensates up to dimension 10,
use the empirical energy scale formula to determine the ideal energy scales of the QCD spectral densities,
and study the
ground state masses and pole residues of the hidden-charm tetraquark states
with quantum numbers $J^{PC}=0^{+\pm}$ and $0^{-\pm}$.
The numerical results of the masses $M_{Z_{0}^{\pm\,,a/b}}$ show
that the charge conjugation
partners have almost degenerate masses for the pseudoscalar tetraquark states,
while there is a considerably large energy gap about $610\,\text{MeV}$  between the masses of the $C=+$
and $C=-1$ scalar tetraquark states, which is especially
interesting.
And the mass predictions of the scalar tetraquark states are larger than
the counterparts of the pseudoscalar tetraquark states.
Moreover, we briefly discuss
the possible decay patterns of the tetraquark states.
Our studies on the tetraquark states can be useful for their searches in
future experiments at facilities such as BESIII, BelleII, LHCb, etc.

\section*{Acknowledgements}
This work is supported by National Natural Science Foundation, Grant Number 11775079.

\section*{Appendix}
The explicit expressions of the QCD spectral densities $\rho^{Z_{0}^{\,t,a/b}}\left(s\right)$ for the scalar and pseudoscalar tetraquark states,
\begin{eqnarray}
\rho_{0}^{t,a}(s)&=&\frac{1}{1024\pi^{6}s}\int_{y_{i}}^{y_{f}}dy\int_{z_{i}}^{1-y}dz\ yz\left(1-y-z\right)^{3}\left(s-\hat{m}_{c}^{2}\right)^{2}\left(7s^2-2\hat{m}_{c}^{2}s-\hat{m}_{c}^{4}\right) \nonumber\\
&&+\frac{tm_{c}^2}{1536\pi^{6}s}\int_{y_{i}}^{y_{f}}dy\int_{z_{i}}^{1-y}dz\, \left(1-y-z\right)^{3}\left(s-\hat{m}_{c}^{2}\right)^{2}\left(7s-\hat{m}_{c}^{2}\right)\ ,
\end{eqnarray}

\begin{eqnarray}
\rho_{0}^{t,b}(s)&=&\frac{1}{1024\pi^{6}s}\int_{y_{i}}^{y_{f}}dy\int_{z_{i}}^{1-y}dz\ yz\left(1-y-z\right)^{3}\left(s-\hat{m}_{c}^{2}\right)^{2}\left(7s^2-2\hat{m}_{c}^{2}s-\hat{m}_{c}^{4}\right)\ ,
\end{eqnarray}

\begin{eqnarray}
\rho_{3}^{t,a}(s)&=&\frac{\left(1+t\right)m_{c}\langle\bar{q}q\rangle}{64\pi^{4}s}\int_{y_{i}}^{y_{f}}dy\int_{z_{i}}^{1-y}dz\, \left(y+z\right)\left(1-y-z\right)\left(s-\hat{m}_{c}^{2}\right)\left(5s-\hat{m}_{c}^{2}\right)\ ,
\end{eqnarray}

\begin{eqnarray}
\rho_{3}^{t,b}(s)&=&\frac{3m_{c}\langle\bar{q}q\rangle}{64\pi^{4}s}\int_{y_{i}}^{y_{f}}dy\int_{z_{i}}^{1-y}dz\, \left(y+z\right)\left(1-y-z\right)\left(s-\hat{m}_{c}^{2}\right)^{2}\ ,
\end{eqnarray}

\begin{eqnarray}
\rho_{4}^{t,a}(s)&=&-\frac{m_{c}^{2}}{768\pi^{4}s}\langle\frac{\alpha_{s}GG}{\pi}\rangle\int_{y_{i}}^{y_{f}}dy\int_{z_{i}}^{1-y}dz\ \left(\frac{z}{y^{2}}+\frac{y}{z^{2}}\right)\left(1-y-z\right)^{3}
\left[\hat{m}_{c}^{2}+\frac{s^2}{3}\delta(s-\hat{m}_{c}^{2})\right]\nonumber\\
&&-\frac{tm_{c}^{4}}{4608\pi^{4}s}\langle\frac{\alpha_{s}GG}{\pi}\rangle\int_{y_{i}}^{y_{f}}dy\int_{z_{i}}^{1-y}dz\ \left(\frac{1}{y^3}+\frac{1}{z^3}\right)\left(1-y-z\right)^{3}\left[1+2s\delta(s-\hat{m}_{c}^{2})\right] \nonumber\\
&&+\frac{1}{512\pi^{4}}\langle\frac{\alpha_{s}GG}{\pi}\rangle\int_{y_{i}}^{y_{f}}dy\int_{z_{i}}^{1-y}dz\ \left(y+z\right)\left(1-y-z\right)^{2}\left(5s-4\hat{m}_{c}^{2}\right) \nonumber\\
&&-\frac{tm_{c}^{2}}{2304\pi^{4}s}\langle\frac{\alpha_{s}GG}{\pi}\rangle\int_{y_{i}}^{y_{f}}dy\int_{z_{i}}^{1-y}dz\ \frac{\left(1-y-z\right)^2}{yz}\left(s-\hat{m}_{c}^{2}\right) \nonumber\\
&&+\frac{tm_{c}^{2}}{13824\pi^{4}s}\langle\frac{\alpha_{s}GG}{\pi}\rangle\int_{y_{i}}^{y_{f}}dy\int_{z_{i}}^{1-y}dz\ \left\{\left[\frac{1}{yz}+9\left(\frac{1}{y^{2}}+\frac{1}{z^{2}}\right)\right]\left(1-y-z\right) \right. \nonumber\\
&&\left.+\frac{27}{2}\left(\frac{1}{y}+\frac{1}{z}\right)\right\}
\left(1-y-z\right)^2\left(3s-\hat{m}_{c}^{2}\right)\ ,
\end{eqnarray}

\begin{eqnarray}
\rho_{4}^{t,b}(s)&=&-\frac{m_{c}^{2}}{768\pi^{4}s}\langle\frac{\alpha_{s}GG}{\pi}\rangle\int_{y_{i}}^{y_{f}}dy\int_{z_{i}}^{1-y}dz\ \left(\frac{y}{z^{2}}+\frac{z}{y^{2}}\right)\left(1-y-z\right)^{3} \left[\hat{m}_{c}^{2}+\frac{s^2}{3}\delta(s-\hat{m}_{c}^{2})\right] \nonumber\\
&&+\frac{1}{512\pi^{4}}\langle\frac{\alpha_{s}GG}{\pi}\rangle\int_{y_{i}}^{y_{f}}dy\int_{z_{i}}^{1-y}dz\ \left(y+z\right)\left(1-y-z\right)^{2}\left(5s-4\hat{m}_{c}^{2}\right) \nonumber\\
&&+\frac{tm_{c}^{2}}{2304\pi^{4}s}\langle\frac{\alpha_{s}GG}{\pi}\rangle\int_{y_{i}}^{y_{f}}dy\int_{z_{i}}^{1-y}dz\  \nonumber\\ &&\left[\frac{\left(1-y-z\right)^{2}}{yz}+2-2\left(\frac{1}{y}+\frac{1}{z}\right)\left(1-y-z\right)\right]\left(s-\hat{m}_{c}^{2}\right) \nonumber\\
&&-\frac{tm_{c}^{2}}{13824\pi^{4}s}\langle\frac{\alpha_{s}GG}{\pi}\rangle\int_{y_{i}}^{y_{f}}dy\int_{z_{i}}^{1-y}dz\  \nonumber\\
&&\left[\frac{\left(1-y-z\right)^{3}}{yz}+6\left(1-y-z\right)-3\left(\frac{1}{y}+\frac{1}{z}\right)\left(1-y-z\right)^{2}\right]
\left(3s-\hat{m}_{c}^{2}\right)\ ,
\end{eqnarray}

\begin{eqnarray}
\rho_{5}^{t,a}(s)&=&\frac{m_{c}\langle\bar{q}g_{s}\sigma Gq\rangle}{128\pi^{4}s}\int_{y_{i}}^{y_{f}}dy\int_{z_{i}}^{1-y}dz\ \left\{t\left[1-2\left(y+z\right)\right]-\left(y+z\right)\right\}\left(3s-\hat{m}_{c}^{2}\right) \nonumber\\
&&+\frac{m_{c}\langle\bar{q}g_{s}\sigma Gq\rangle}{128\pi^{4}s}\int_{y_{i}}^{y_{f}}dy\int_{z_{i}}^{1-y}dz\ \left(\frac{y}{z}+\frac{z}{y}\right)\left(1-y-z\right)\hat{m}_{c}^{2} \nonumber\\
&&-\frac{tm_{c}\langle\bar{q}g_{s}\sigma Gq\rangle}{384\pi^{4}s}\int_{y_{i}}^{y_{f}}dy\int_{z_{i}}^{1-y}dz\ \left(\frac{y}{z}+\frac{z}{y}\right)\left(1-y-z\right)\left(s-\hat{m}_{c}^{2}\right)\ ,
\end{eqnarray}

\begin{eqnarray}
\rho_{5}^{t,b}(s)&=&+\frac{m_{c}\langle\bar{q}g_{s}\sigma Gq\rangle}{384\pi^{4}s}\int_{y_{i}}^{y_{f}}dy\int_{z_{i}}^{1-y}dz\ \left[t\left(\frac{y}{z}+\frac{z}{y}\right)\left(1-y-z\right)-\left(t+9\right)\left(y+z\right)\right]\left(s-\hat{m}_{c}^{2}\right) \nonumber\\
&&-\frac{m_{c}\langle\bar{q}g_{s}\sigma Gq\rangle}{128\pi^{4}s}\int_{y_{i}}^{y_{f}}dy\int_{z_{i}}^{1-y}dz\ \left(\frac{y}{z}+\frac{z}{y}\right)\left(1-y-z\right)\hat{m}_{c}^{2} \ ,
\end{eqnarray}

\begin{eqnarray}
\rho_{6}^{t,a}(s)&=&\frac{m_{c}^{2}\langle\bar{q}q\rangle^2}{12\pi^{2}s}\int_{y_{i}}^{y_{f}}dy+\frac{t\langle\bar{q}q\rangle^2}{24\pi^{2}s} \int_{y_{i}}^{y_{f}}dy\ y\left(1-y\right)\left(3s-\tilde{m}_{c}^{2}\right) \nonumber\\
&&+\frac{tm_c^{2}g_{s}^{2}\langle\bar{q}q\rangle^{2}}{5184\pi^{4}s}\int_{y_{i}}^{y_{f}}dy\int_{z_{i}}^{1-y}dz\ \left[4-5\left(\frac{1}{y}+\frac{1}{z}\right)\left(1-y-z\right)\right]\left[1+2s\delta(s-\hat{m}_{c}^{2})\right] \nonumber\\
&&-\frac{g_{s}^{2}\langle\bar{q}q\rangle^{2}}{2592\pi^{4}s}\int_{y_{i}}^{y_{f}}dy\int_{z_{i}}^{1-y}dz\
\left\{\left[\left(y+z\right)+14\left(\frac{y}{z}+\frac{z}{y}\right)\right]\left(1-y-z\right)-12yz\right\}\hat{m}_{c}^{2} \nonumber\\
&&+\frac{g_{s}^{2}\langle\bar{q}q\rangle^{2}}{864\pi^{4}}\int_{y_{i}}^{y_{f}}dy\int_{z_{i}}^{1-y}dz\ \left[3\left(\frac{y}{z}+\frac{z}{y}\right)-4\left(y+z\right)\right]\left(1-y-z\right) \nonumber\\
&&-\frac{g_{s}^{2}\langle\bar{q}q\rangle^{2}}{3888\pi^{4}}\int_{y_{i}}^{y_{f}}dy\int_{z_{i}}^{1-y}dz\ \left[5\left(y+z\right)\left(1-y-z\right)-6yz\right]s\delta(s-\hat{m}_{c}^{2}) \nonumber\\
&&+\frac{m_c^{2}g_{s}^{2}\langle\bar{q}q\rangle^{2}}{7776\pi^{4}s}\int_{y_{i}}^{y_{f}}dy\int_{z_{i}}^{1-y}dz\ \left(\frac{y}{z^2}+\frac{z}{y^2}\right)\left(1-y-z\right)\left[23-5s\delta(s-\hat{m}_{c}^{2})\right]
\ ,
\end{eqnarray}

\begin{eqnarray}
\rho_{6}^{t,b}(s)&=& -\frac{m_{c}^{2}\langle\bar{q}q\rangle^2}{12\pi^{2}s}\int_{y_{i}}^{y_{f}}dy
+\frac{g_{s}^{2}\langle\bar{q}q\rangle^2}{864\pi^{4}}\int_{y_{i}}^{y_{f}}dy\int_{z_{i}}^{1-y}dz\ \left[3\left(\frac{y}{z}+\frac{z}{y}\right)-4\left(y+z\right)\right]\left(1-y-z\right) \nonumber\\
&&-\frac{g_{s}^{2}\langle\bar{q}q\rangle^2}{2592\pi^{4}s}\int_{y_{i}}^{y_{f}}dy\int_{z_{i}}^{1-y}dz\ \left\{\left[\left(y+z\right)+14\left(\frac{y}{z}+\frac{z}{y}\right)\right]\left(1-y-z\right)-12yz\right\}
\hat{m}_{c}^{2} \nonumber\\
&&-\frac{g_{s}^{2}\langle\bar{q}q\rangle^2}{3888\pi^{4}}\int_{y_{i}}^{y_{f}}dy\int_{z_{i}}^{1-y}dz\ \left[5\left(y+z\right)\left(1-y-z\right)-6yz\right]
s\delta(s-\hat{m}_{c}^{2}) \nonumber\\
&&+\frac{m_{c}^{2}g_{s}^{2}\langle\bar{q}q\rangle^2}{7776\pi^{4}s}\int_{y_{i}}^{y_{f}}dy\int_{z_{i}}^{1-y}dz\ \left(\frac{y}{z^2}+\frac{z}{y^2}\right)\left(1-y-z\right)\left[23-5s\delta(s-\hat{m}_{c}^{2})\right]
\ ,
\end{eqnarray}

\begin{eqnarray}
\rho_{7}^{t,a}(s)&=&\frac{m_{c}\langle\bar{q}q\rangle}{192\pi^{2}s}\langle\frac{\alpha_{s}GG}{\pi}\rangle \int_{y_{i}}^{y_{f}}dy\int_{z_{i}}^{1-y}dz
 \left[\left(1+t\right)\left(\frac{y}{z^{2}}+\frac{z}{y^{2}}\right)\left(1-y-z\right)+\frac{t}{2}\left(\frac{y}{z}+\frac{z}{y}\right)+3\right] \nonumber\\
&&+\frac{m_{c}\langle\bar{q}q\rangle}{96\pi^{2}}\langle\frac{\alpha_{s}GG}{\pi}\rangle \int_{y_{i}}^{y_{f}}dy\int_{z_{i}}^{1-y}dz \left[\left(1+t\right)\left(\frac{y}{z^{2}}+\frac{z}{y^{2}}\right)\left(1-y-z\right)+\frac{t}{2}\left(\frac{y}{z}+\frac{z}{y}\right)+1\right]
\nonumber\\
&&\delta(s-\hat{m}_{c}^{2})
-\frac{\left(1+t\right)m_{c}^{3}\langle\bar{q}q\rangle}{576\pi^{2}}\langle\frac{\alpha_{s}GG}{\pi}\rangle \int_{y_{i}}^{y_{f}}dy\int_{z_{i}}^{1-y}dz \left(\frac{1}{y^{2}}+\frac{1}{z^{2}}+\frac{y}{z^3}+\frac{z}{y^3}\right)(1-y-z) \nonumber\\
&&\left(\frac{1}{s}+\frac{2}{T^2}\right)\delta(s-\hat{m}_{c}^{2}) +\frac{\left(1+t\right)m_{c}\langle\bar{q}q\rangle}{1152\pi^{2}s}\langle\frac{\alpha_{s}GG}{\pi}\rangle\int_{y_{i}}^{y_{f}}dy\ \left[1+2s\delta(s-\tilde{m}_{c}^{2})\right]
\ ,
\end{eqnarray}

\begin{eqnarray}
\rho_{7}^{t,b}(s)&=&\frac{m_{c}\langle\bar{q}q\rangle}{576\pi^{2}s}\langle\frac{\alpha_{s}GG}{\pi}\rangle\int_{y_{i}}^{y_{f}}dy\int_{z_{i}}^{1-y}dz \left\{\left[9\left(\frac{y}{z^{2}}+\frac{z}{y^{2}}\right)-t\left(\frac{1}{y}+\frac{1}{z}\right)\right]\left(1-y-z\right)+(2t+3)\right\} \nonumber\\
&&-\frac{m_{c}^{3}\langle\bar{q}q\rangle}{192\pi^{2}s}\langle\frac{\alpha_{s}GG}{\pi}\rangle\int_{y_{i}}^{y_{f}}dy\int_{z_{i}}^{1-y}dz \left(\frac{1}{y^2}+\frac{1}{z^2}+\frac{y}{z^3}+\frac{z}{y^3}\right)\left(1-y-z\right)\delta(s-\hat{m}_{c}^{2})
 \nonumber\\
&&+\frac{m_{c}\langle\bar{q}q\rangle}{384\pi^{2}s}\langle\frac{\alpha_{s}GG}{\pi}\rangle\int_{y_{i}}^{y_{f}}dy
\ ,
\end{eqnarray}

\begin{eqnarray}
\rho_{8}^{t,a}(s)&=&-\frac{\langle\bar{q}q\rangle\langle\bar{q}g_{s}\sigma Gq\rangle}{24\pi^{2}}\int_{y_{i}}^{y_{f}}dy \left(\frac{2m_{c}^{2}}{s}+\frac{m_{c}^{2}}{T^{2}}-\frac{1}{4}\right)\delta(s-\tilde{m}_{c}^{2}) \nonumber\\
&&-\frac{7t\langle\bar{q}q\rangle\langle\bar{q}g_{s}\sigma Gq\rangle}{48\pi^{2}}\int_{y_{i}}^{y_{f}}dy\ y\left(1-y\right)\left(1+\frac{2s}{7T^2}\right)\delta(s-\tilde{m}_{c}^{2}) \nonumber\\
&&+\frac{t\langle\bar{q}q\rangle\langle\bar{q}g_{s}\sigma Gq\rangle}{192\pi^{2}s}\int_{y_{i}}^{y_{f}}dy\ \left[1+2s\delta(s-\tilde{m}_{c}^{2})-12y\left(1-y\right)\right]
\ ,
\end{eqnarray}

\begin{eqnarray}
\rho_{8}^{t,b}(s)&=&\frac{\langle\bar{q}q\rangle\langle\bar{q}g_{s}\sigma Gq\rangle}{24\pi^{2}}\int_{y_{i}}^{y_{f}}dy\ \left[\frac{2m_{c}^{2}}{s}+\frac{m_{c}^{2}}{T^{2}}-\frac{1}{4}\right]\delta(s-\tilde{m}_{c}^{2}) \nonumber\\
&&-\frac{t\langle\bar{q}q\rangle\langle\bar{q}g_{s}\sigma Gq\rangle}{288\pi^{2}s}\int_{y_{i}}^{y_{f}}dy\ \ ,
\end{eqnarray}

\begin{eqnarray}
\rho_{10}^{t,a}(s)&=&t\left(\frac{\langle\bar{q}g_{s}\sigma Gq\rangle^2}{32\pi^{2}}+\frac{\langle\bar{q}q\rangle^2}{36}\langle\frac{\alpha_{s}GG}{\pi}\rangle\right)\int_{y_{i}}^{y_{f}}dy\,y\left(1-y\right)
\left(\frac{1}{s}+\frac{3}{2T^2}+\frac{3s}{4T^4}+\frac{s^2}{6T^6}\right) \delta(s-\tilde{m}_{c}^{2}) \nonumber\\
&&+\left(\frac{m_{c}^{2}\langle\bar{q}g_{s}\sigma Gq\rangle^2}{32\pi^{2}s^2}
+\frac{m_{c}^{2}\langle\bar{q}q\rangle^2}{36s^2}\langle\frac{\alpha_{s}GG}{\pi}\rangle\right)\int_{y_{i}}^{y_{f}}dy\ \left(1+\frac{s}{T^2}+\frac{s^2}{2T^4}+\frac{s^3}{6T^6}\right)\delta(s-\tilde{m}_{c}^{2}) \nonumber\\
&&-\frac{\langle\bar{q}g_{s}\sigma Gq\rangle^2}{192\pi^{2}s}\int_{y_{i}}^{y_{f}}dy\,
\left(1+\frac{s}{T^2}+\frac{s^2}{2T^4}\right) \delta(s-\tilde{m}_{c}^{2}) \nonumber\\
&&-\frac{t\langle\bar{q}g_{s}\sigma Gq\rangle^2}{384\pi^{2}s}\int_{y_{i}}^{y_{f}}dy\ \left(1+\frac{3s}{2T^2}+\frac{s^2}{T^4}\right)\delta(s-\tilde{m}_{c}^{2}) \nonumber\\
&&+\frac{m_{c}^{2}\langle\bar{q}q\rangle^2}{72s^2}\langle\frac{\alpha_{s}GG}{\pi}\rangle\int_{y_{i}}^{y_{f}}dy\ \left[\frac{1}{y^2}+\frac{1}{\left(1-y\right)^2}\right]
\left(1+\frac{s}{T^2}\right)\delta(s-\tilde{m}_{c}^{2}) \nonumber\\
&&-\frac{m_{c}^{4}\langle\bar{q}q\rangle^2}{108s^3}\langle\frac{\alpha_{s}GG}{\pi}\rangle\int_{y_{i}}^{y_{f}}dy\ \left[\frac{1}{y^3}+\frac{1}{\left(1-y\right)^3}\right]
\left(1+\frac{s}{T^2}+\frac{s^2}{2T^4}\right) \delta(s-\tilde{m}_{c}^{2}) \nonumber\\
&&-\frac{tm_{c}^{2}\langle\bar{q}q\rangle^2}{432s^2}\langle\frac{\alpha_{s}GG}{\pi}\rangle\int_{y_{i}}^{y_{f}}dy\ \left[\frac{y}{\left(1-y\right)^2}+\frac{1-y}{y^2}\right]
\left(1+\frac{s}{T^2}+\frac{2s^2}{T^4}\right) \delta(s-\tilde{m}_{c}^{2})
 \ ,
\end{eqnarray}

\begin{eqnarray}
\rho_{10}^{t,b}(s)&=&-\left(\frac{m_{c}^{2}\langle\bar{q}g_{s}\sigma Gq\rangle^2}{32\pi^{2}s^2}+\frac{m_{c}^{2}\langle\bar{q}q\rangle^2}{36s^2}\langle\frac{\alpha_{s}GG}{\pi}\rangle\right)\int_{y_{i}}^{y_{f}}dy\ \left(1+\frac{s}{T^2}+\frac{s^2}{2T^4}+\frac{s^3}{6T^6}\right) \delta(s-\tilde{m}_{c}^{2}) \nonumber\\
&&-\frac{m_{c}^{2}\langle\bar{q}q\rangle^2}{72s^2}\langle\frac{\alpha_{s}GG}{\pi}\rangle\int_{y_{i}}^{y_{f}}dy\ \left[\frac{1}{y^2}+\frac{1}{\left(1-y\right)^2}\right]\left(1+\frac{s}{T^2}\right) \delta(s-\tilde{m}_{c}^{2}) \nonumber\\
&&+\frac{m_{c}^{4}\langle\bar{q}q\rangle^2}{108s^3}\langle\frac{\alpha_{s}GG}{\pi}\rangle\int_{y_{i}}^{y_{f}}dy\ \left[\frac{1}{y^3}+\frac{1}{\left(1-y\right)^3}\right] \left(1+\frac{s}{T^2}+\frac{s^2}{2T^4}\right) \delta(s-\tilde{m}_{c}^{2}) \nonumber\\
&&+\frac{t\langle\bar{q}q\rangle^2}{432s}\langle\frac{\alpha_{s}GG}{\pi}\rangle\int_{y_{i}}^{y_{f}}dy\ \delta(s-\tilde{m}_{c}^{2}) +\frac{t\langle\bar{q}g_{s}\sigma Gq\rangle^2}{256\pi^{2}s}\int_{y_{i}}^{y_{f}}dy\ \left(1+\frac{2s}{9T^2}\right)\delta(s-\tilde{m}_{c}^{2}) \nonumber\\
&&+\frac{\langle\bar{q}g_{s}\sigma Gq\rangle^2}{192\pi^{2}s}\int_{y_{i}}^{y_{f}}dy\,
\left(1+\frac{s}{T^2}+\frac{s^2}{2T^4}\right) \delta(s-\tilde{m}_{c}^{2})
\ ,
\end{eqnarray}
where $y_{f}=\frac{1+\sqrt{1-4m_{c}^{2}/s}}{2}$, $y_{i}=\frac{1-\sqrt{1-4m_{c}^{2}/s}}{2}$, $z_{i}=\frac{ym_{c}^{2}}{ys-m_{c}^{2}}$, $\hat{m}_{c}^{2}=\frac{(y+z)m_{c}^{2}}{yz}$, $\tilde{m}_{c}^{2}=\frac{m_{c}^{2}}{y(1-y)}$, $\int_{y_{i}}^{y_{f}}dy\rightarrow\int_{0}^{1}$, $\int_{z_{i}}^{1-y}dz\rightarrow\int_{0}^{1-y}dz$, when the $\delta$ functions $\delta(s-\hat{m}_{c}^{2})$ and $\delta(s-\tilde{m}_{c}^{2})$ appear.


\begin{thebibliography}{99}

\bibitem{X(3872)} S. K. Choi, et al., Phys. Rev. Lett. {\bf 91}, 262001 (2003).

\bibitem{XYZ} C. Patrignani, et al., Chin. Phys. C {\bf  40}, 100001 (2016).

\bibitem{Pc} R. Aaij, et al., Phys. Rev. Lett. {\bf 115}, 072001 (2015).

\bibitem{Quark-Model} S. Godfrey and N. Isgur, Phys. Rev. D {\bf  32}, 189 (1985).

\bibitem{Hadron-Molecules} M. Voloshin and L. Okun, JETP Lett. {\bf 23}, 333 (1976).

\bibitem{Molecular-Charmonium} A. De Rujula, H. Georgi and S. L. Glashow, Phys. Rev. Lett. {\bf 38}, 317 (1977).

\bibitem{Diquarks} M. Anselmino, E. Predazzi, S. Ekelin, S. Fredriksson and D. Lichtenberg, Rev. Mod. Phys. {\bf 65}, 1199 (1993).

\bibitem{tetraquark-1} L. Maiani, F. Piccinini, A. Polosa and V. Riquer, Phys. Rev. D {\bf  71}, 014028 (2005).

\bibitem{tetraquark-2} R. D. Matheus, S. Narison, M. Nielsen and J. M. Richard, Phys. Rev. D {\bf  75}, 014005 (2007).

\bibitem{tetraquark-3} F. S. Navarra, M. Nielsen and S. H. Lee, Phys. Lett. B {\bf  649}, 166 (2007).

\bibitem{tetraquark-4} S. H. Lee, A. Mihara, F. S. Navarra and M. Nielsen, Phys. Lett. B {\bf  661}, 28 (2008).

\bibitem{tetraquark-5} R. M. Albuquerque and M. Nielsen, Nucl. Phys. A {\bf  815}, 53 (2009).

\bibitem{tetraquark-6} Z. G. Wang, Eur. Phys. J. C {\bf  59}, 675 (2009).

\bibitem{tetraquark-7} Z. G. Wang, Eur. Phys. J. C {\bf  62}, 375 (2009).

\bibitem{tetraquark-8} W. Chen and S. L. Zhu, Phys. Rev. D {\bf  83}, 034010 (2011).

\bibitem{tetraquark-9} J. R. Zhang and M. Q. Huang, Phys. Rev. D {\bf  83}, 036005 (2011).

\bibitem{tetraquark-10} C. F. Qiao and L. Tang, Eur. Phys. J. C {\bf  74}, 2810 (2014).

\bibitem{tetraquark-11} J. R. Zhang, J. L. Zou and J. Y. Wu, Chin. Phys. C {\bf  42}, 043101 (2018).

\bibitem{tetraquark-12} H. Sundu, S. S. Agaev and K. Azizi,  Phys. Rev. D {\bf  97}, 054001 (2018).




\bibitem{Pentaquark-1} R. L. Jaffe and F. Wilczek, Phys. Rev. Lett. {\bf 91}, 232003 (2003).

\bibitem{Pentaquark-2} R. Lebed, Phys. Lett. B {\bf  749}, 454 (2015).

\bibitem{Pentaquark-3} L. Maiani, A. Polosa and V. Riquer, Phys. Lett. B {\bf  749}, 289 (2015).

\bibitem{Pentaquark-4} V. V. Anisovich, M. A. Matveev, J. Nyiri, A. V. Sarantsev and A. N. Semenova, arXiv:1507.07652.

\bibitem{Pentaquark-5} H. X. Chen, W. Chen, X. Liu, T. G. Steele and S. L. Zhu, Phys. Rev. Lett. {\bf 115}, 172001 (2015).

\bibitem{Pentaquark-6} G. N. Li, M. He and X. G. He, JHEP {\bf 12}, 128 (2015).

\bibitem{Pentaquark-7} Z. G. Wang, Nucl. Phys. B {\bf  913}, 163 (2016).

\bibitem{Pentaquark-8} Z. G. Wang and T. Huang, Eur. Phys. J. C {\bf  76}, 43 (2016).

\bibitem{Pentaquark-9} Z. G. Wang, Eur. Phys. J. C {\bf  76}, 70 (2016).

\bibitem{Pentaquark-10} S. S. Agaev, K. Azizi and H. Sundu, Eur. Phys. J. C {\bf  77}, 321 (2017).


\bibitem{Quarkonium} M. Voloshin, Prog. Part. Nucl. Phys. {\bf 61}, 455 (2008).

\bibitem{Hadro-Charmonium} S. Dubynskiy, M. Voloshin, Phys. Lett. B {\bf  666}, 344 (2008).


\bibitem{review-1} R. F. Lebed, R. E. Mitchell and E. S. Swanson, Prog. Part. Nucl. Phys. {\bf 93}, 143 (2017).

\bibitem{review-2} H. X. Chen, W. Chen, X. Liu and S. L. Zhu, Phys. Rept. {\bf 639}, 1 (2016).

\bibitem{article-Di} Z. Y. Di, Z. G. Wang, J. X. Zhang and G. L. Yu, Commun. Theor. Phys. {\bf 69}, 191 (2018).

\bibitem{vector} Z. G. Wang, Eur. Phys. J. C {\bf  74}, 2874 (2014).


\bibitem{axialvector} Z. G. Wang and T. Huang, Phys. Rev. D {\bf  89}, 054019 (2014).

\bibitem{meson-loop} Z. G. Wang, Eur. Phys. J. C {\bf  77}, 174 (2017).

\bibitem{PRT85} L. J. Reinders, H. Rubinstein and S. Yazaki, Phys. Rept. {\bf 127}, 1 (1985).


\bibitem{NPB147} M. A. Shifman, A. I. Vainshtein and V. I. Zakharov, Nucl. Phys. B {\bf  147}, 385 (1979).

\bibitem{175} P. Colangelo and A. Khodjamirian, At the Frontier of Particle Physics: Handbook of QCD, Vol. {\bf  3}, ed. M. Shifman (World Scientific, 2001), pp. 1495.

\bibitem{TP63-1} L. Maiani, F. Piccinini, A. D. Polosa and V. Riquer, Phys. Rev. D {\bf  89}, 114010 (2014).

\bibitem{TP63-2} M. Nielsen and F. S. Navarra, Mod. Phys. Lett. A {\bf  29}, 1430005 (2014).

\bibitem{TP63-3} Z. G. Wang, Commun. Theor. Phys. {\bf 63}, 325 (2015).

\bibitem{TP63-4} S. S. Agaev, K. Azizi and H. Sundu, Phys. Rev. D {\bf  96}, 034026 (2017).


\bibitem{EPJC77} Z. G. Wang, Eur. Phys. J. C {\bf  77}, 78 (2017).

\bibitem{energy-scale-1} Z. G. Wang, Commun. Theor. Phys. {\bf 63}, 466 (2015).

\bibitem{energy-scale-2} Z. G. Wang and Y. F. Tian, Int. J. Mod. Phys. A {\bf  30}, 1550004 (2015).

\bibitem{energy-scale-3} Z. G. Wang and T. Huang, Eur. Phys. J. C {\bf  74}, 2891 (2014).

\bibitem{energy-scale-4} Z. G. Wang, Eur. Phys. J. C {\bf  74}, 2963 (2014).

\bibitem{lowest-tetraquark} Z. G. Wang, Mod. Phys. Lett. A {\bf  29}, 1450207 (2014).



\end{thebibliography}
\end{document}